\documentclass[journal,]{IEEEtran}

\usepackage{enumerate}
\usepackage{cite}
\usepackage{tikz}
\usepackage{xcolor}
\graphicspath{{figures/}}
\usepackage{algorithm }
\usepackage{algorithmic}

\usepackage{makecell}
\usepackage{bm}
\usepackage{amsmath,amsfonts,amssymb,amsthm}
\usepackage{subfigure}
\usepackage{threeparttable}
\usepackage{booktabs}
\usepackage{bm}
\usepackage{mathrsfs}
\usepackage{flushend}
\usepackage{color}

\usepackage{url}
\usepackage{hyperref}

\usepackage{footnote}
\makesavenoteenv{tabular}

\begin{document}

\begin{table*}
\centering
\begin{tabular}{p{18cm} }
\\ This paper has been accepted for publication in IEEE Internet of Things Journal\\  \\ 

\\ DOI: 10.1109/JIOT.2018.2889303\\  \\ 

\\ IEEE Explore \url{https://ieeexplore.ieee.org/document/8586939} \\ \\ 

\\ \\Please cite the paper as: \\

\\ Y. Li, Z. He, Z. Gao, Y. Zhuang, C. Shi and N. El-Sheimy, "Toward Robust Crowdsourcing-Based Localization: A Fingerprinting Accuracy Indicator Enhanced Wireless/Magnetic/Inertial Integration Approach," in IEEE Internet of Things Journal, vol. 6, no. 2, pp. 3585-3600, April 2019. \\

\\ \\ bibtex:  \\

\\ @article\{LiY-IOTJ-Loc, \\
  author=\{Y. \{Li\} and Z. \{He\} and Z. \{Gao\} and Y. \{Zhuang\} and C. \{Shi\} and N. \{El-Sheimy\}\},\\
  journal=\{IEEE Internet of Things Journal\}, \\
 title=\{Toward Robust Crowdsourcing-Based Localization: A Fingerprinting Accuracy Indicator Enhanced Wireless/Magnetic/Inertial Integration Approach\},\\ 
  year=\{2019\},\\
  volume=\{6\},\\
  number=\{2\},\\
  pages=\{3585-3600\},\\
  doi=\{10.1109/JIOT.2018.2889303\},\\
  ISSN=\{2327-4662\},\\
  month=\{April\}\,\\
  \}\\

 \end{tabular}
\end{table*}

\clearpage

\title{Towards Robust Crowdsourcing-Based Localization: A Fingerprinting Accuracy Indicator Enhanced Wireless/Magnetic/Inertial Integration Approach \thanks{
}}
\author{You~Li,
  Zhe~He,
  Zhouzheng Gao,
  Yuan Zhuang,
  Chuang Shi,
  and Naser El-Sheimy
   \thanks{You Li, Zhe He, and Naser El-Sheimy are with Department of Geomatics Engineering, University of Calgary, 2500 University Dr NW, Calgary, AB T2N 1N4. Yuan Zhuang is with the State Key Laboratory of Surveying, Mapping and Remote Sensing, Wuhan University, 129 Luoyu Road, Wuhan, China, 430079, zhy.0908@gmail.com. Zhouzheng Gao is with School of Land Science and Technology, China University of Geosciences (Beijing), 29 Xueyuan Road, Beijing, China, 100083. Chuang Shi is with the School of Electronic and Information Engineering, Beihang University, 37 Xueyuan Road, Beijing, China, 100083. This paper is partly supported by the Natural Sciences and Engineering Research Council of Canada (NSERC) and the National Natural Science Foundation of China (NSFC) (Grant No. 61771135, 41804027).}
}

\markboth{}%
         {Shell \MakeLowercase{\textit{et al.}}: Bare Demo of IEEEtran.cls for Journals}

         \maketitle

         \begin{abstract}
         The next-generation internet of things (IoT) systems have an increasingly demand on intelligent localization which can scale with big data without human perception. Thus, traditional localization solutions without accuracy metric will greatly limit vast applications. Crowdsourcing-based localization has been proven to be effective for mass-market location-based IoT applications. This paper proposes an enhanced crowdsourcing-based localization method by integrating inertial, wireless, and magnetic sensors. Both wireless and magnetic fingerprinting accuracy are predicted in real time through the introduction of fingerprinting accuracy indicators (FAI) from three levels (i.e., signal, geometry, and database). The advantages and limitations of these FAI factors and their performances on predicting location errors and outliers are investigated. Furthermore, the FAI-enhanced extended Kalman filter (EKF) is proposed, which improved the dead-reckoning (DR)/WiFi, DR/Magnetic, and DR/WiFi/Magnetic integrated localization accuracy by 30.2 \%, 19.4 \%, and 29.0 \%, and reduced the maximum location errors by 41.2 \%, 28.4 \%, and 44.2 \%, respectively. These outcomes confirm the effectiveness of the FAI-enhanced EKF on improving both accuracy and reliability of multi-sensor integrated localization using crowdsourced data.        
         \end{abstract}
		 
         \begin{IEEEkeywords}
         Crowdsourcing; internet of things; location based services; inertial sensor; received signal strength; magnetic matching; fingerprinting. 
         \end{IEEEkeywords}

         \IEEEpeerreviewmaketitle

         \section{Introduction}
         \subsection{Problem Statement}
         \IEEEPARstart{T}{he} advent of internet of things (IoT) era is making it possible to ``locate" and ``connect" anywhere because of the promotion of public wireless (e.g., low-power wide-area network (LPWAN) \cite{UPINLBS}, cellulars, wireless local area network (WiFi), and Bluetooth) base stations and the availability of sensors in smart IoT devices. Crowdsourcing using data from existing public infrastructures is a trend for low-cost mass-market IoT applications \cite{Xiang2017}. However, traditional indoor localization performance in mass-market applications are difficult to predict: the localization accuracy may reach as accurate as one meter in environments that have a strong signal geometry, or be degraded to tens of meters. The unpredictability of positioning accuracy has limited the \textit{scalability} of localization in IoT applications. Although there are extensive research that have made significant improvements on localization under specific scenarios, it is difficult to maintain localization accuracy ubiquitously due to the complexity of daily-life scenarios, the diversity in low-cost devices, and the requirement for system deployments.
         
         An approach for alleviating the scalability issue is to predict the localization accuracy (or uncertainty) in real time, so as to adaptively adjust the location solutions for various scenarios. Till today, most existing indoor localization works focus on improving localization techniques, while there is a lack of investigation of indicator metric regarding the localization accuracy, especially fingerprinting accuracy. Therefore, the main objective of this paper is to provide a reference on predicting fingerprinting accuracy and using it to improve localization.
                  
         To begin with, the related state-of-the-art techniques, such as those for low-cost indoor localization, crowdsourcing-based localization, and localization accuracy prediction, are reviewed, followed by the contributions of this paper.
          
        \subsection{Low-Cost Indoor Localization Techniques}
        There are generally three types of indoor localization techniques: the geometrical methods, fingerprinting, and dead-reckoning (DR) based algorithms. Typical geometrical methods include multilateration that uses device-access point (AP) ranges and triangulation that is based on angle-of-arrival measurements or ranges. Ranging measurements can either be calculated from received signal strength (RSS) by using path-loss models, or be determined by time-of-arrival measurements. Time-of-arrival and angle-of-arrival measurements require specific hardware for precise time synchronization and phase synchronization, respectively, while RSS is the mainstream wireless signal that is supported by the majority of IoT devices. For RSS-based methods, multilateration has advantages such as a small database, the capability of positioning in areas beyond the database, the extrapolation capability for uncharted areas, and most importantly, the feasibility to obtain real-time positioning accuracy. However, for indoor scenarios, the multilateration performance may be degraded by non-line-of-sight conditions. Thus, it is difficult for multilateration to provide accuracy that is competitive to fingerprinting in indoor areas. Therefore, fingerprinting and DR are the most widely used methods for low-cost indoor localization. 
        
        \subsubsection{Fingerprinting}
        \label{sub-fp}
        Fingerprinting has been proven to be effective for localization using signals such as RSS and magnetic data \cite{LiY-PhD}. Fingerprinting consists of two steps: training and testing. For training, [feature vector, location] fingerprints at multiple reference points (RPs) are used to generate a database. For testing, RPs with the closest features to the measured data are selected to compute the device location. To determine the closest features, the likelihood (or weight) for each fingerprint can be computed through the deterministic (e.g., nearest neighbors \cite{Fu2018}), probabilistic (e.g., Gaussian distribution \cite{Haeberlen2004} and Gauss process \cite{HeZ2018}), and machine-learning (e.g., random forest \cite{Guo2018} and neural networks \cite{FangS2008}) methods. A key challenge for fingerprinting is the inconsistency between database and real-time measurements. Examples of factors that lead to such inconsistency include a) database timeliness; b) signal fluctuations and interference such as reflection, multipath, and disturbances by human body; c) device diversity; and d) device orientation and motion mode. For a), there are database-updating methods through techniques such as crowdsourcing \cite{ZhangP-Sensors} and simultaneous localization and mapping (SLAM) \cite{Gentner2016}. For b), possible solutions include calibration-based methods \cite{Lee2015}, and calibration-free methods such as differential RSS \cite{Hossain2013}. For c), practical methods include RSS filtering \cite{Xue2017}, the selective use of APs \cite{Liang2015}, and the use of advanced models for multipath \cite{Gentner2016}, non-line-of-sight \cite{HeZ2014}, fading \cite{Akyildiz2009}, and channel diversity \cite{Zanella2014}. Finally, for d), there are methods by orientation aiding \cite{Sanchez2012}, database selection \cite{King2006}, and orientation compensation \cite{Fang2015}. In general, improving fingerprinting performance in specific scenarios has been well researched. 
           		
   		\subsubsection{Dead-Reckoning}
   		\label{sub-dr}
   		Low-cost motion sensor (e.g., inertial sensor and magnetometer) based DR can provide autonomous indoor/outdoor localization solutions \cite{YuN2018}. However, obtaining a long-term accurate DR solution is challenging due to factors such as a) the requirement for heading and position initialization; b) the existence of sensor errors and drifts, and c) the variety of device motions that brings in misalignment angles between vehicle (e.g., human body) and device. For a), wireless position and magnetic heading are commonly used for coarse location and heading initialization, while a filter (e.g., Kalman filter or particle filter) is used to update position and heading gradually. Thus, the accuracy for both initial and following-up location updates is important. For b), real-time calibration for gyros \cite{LiY2015} and magnetometers \cite{Wahdan2014} can be used. For c), methods such as principal component analysis \cite{DengZ2015} are utilized for misalignment estimation. Meanwhile, the key for using low-cost DR is to extract constraints, such as the pedestrian velocity model \cite{LiY-2018-access}, multi-device constraints \cite{LanH-2015}, and the building heading constraint \cite{Abdulrahim2010} for indoor pedestrian localization. Short-term DR solutions can bridge the outages for external localization signals, provide smoother and more reliable solutions when integrated with other techniques, and aid the profile matching of wireless \cite{LiY2016} and magnetic signals \cite{LiY-Mag}.
   		 
   		\subsubsection{Multi-Sensor Integration}
   		Due to the complexity of daily-life scenarios, it is challenging to provide low-cost but high-performance localization solutions by using a standalone technology. Because of the complementary characteristics among various technologies, multi-sensor integration is widely adopted. The data from motion sensors are commonly used to build the system model and provide a short-term prediction, while technologies such as wireless and magnetic fingerprinting provide the updates. Kalman filter \cite{LiY2017} and particle filter \cite{Pak2015} are widely used techniques for information fusion. The majority of literature integrates multi-sensor information through a loosely-coupled way, while others apply a tightly-coupled approach \cite{ZhuangY2018}. The key in multi-sensor integration is to take advantage of the merits of each single technique according to specific scenarios. This is commonly achieved by setting and tuning the weight for each technique in the filter, which is described in Subsection \ref{sub-review-fai}.  
   		
         \subsection{Crowdsourcing-Based Localization}
         While millions of users and devices are connected, massive amount of data are collected in real time. Accordingly, ``evolution" is needed for the traditional localization modes that consist of separate training and testing steps. Under the crowdsourcing framework, location users become also location providers, as their localization solutions may be used to update the database. There are crowdsourcing-based localization approaches \cite{Zhuang-EL} that estimate AP locations and path-loss model parameters (PPs) with crowdsourced data, and SLAM-based methods \cite{Gentner2016} that update databases while localizing. For crowdsourcing or SLAM based methods, a challenge is to obtain robust path predictions, updates, and constraints for optimization of the graph that contains device and AP parameters. DR is widely used to provide path predictions, while the fingerprinting solution is utilized to provide position updates or constraints. Therefore, the challenges for DR in Subsection \ref{sub-dr} and fingerprinting in Subsection \ref{sub-fp} are also issues to solve for crowdsourcing-based localization. Meanwhile, the selection of robust sensor data and fingerprints is key to a reliable crowdsourcing system. The research \cite{ZhangP2018} proposes a crowdsourced sensor data quality assessment framework. In this paper, the location accuracy indicator is further introduced, which can be used to predict the quality of location updates before using them for multi-sensor integration or database updating, and in turn enhance the reliability of crowdsourcing solutions. Existing works about localization accuracy indicators are discussed in the next subsection. 
                   
         \subsection{Localization Accuracy Prediction}
         \label{sub-review-fai}
         The multi-sensor integration performance is highly dependent on the setting of parameters, such as the initial state covariance matrix ($\textbf{P}$), the system noise covariance matrix ($\textbf{Q}$), and the measurement noise covariance matrix ($\textbf{R}$). Especially, improper setting of $\textbf{Q}$ or $\textbf{R}$ may lead to degradation \cite{Mehra1970} or even divergence \cite{Fitzgerald1971} of estimation. Since the Kalman filter system model is constructed by self-contained motion models, elements in $\textbf{Q}$ are set according to the stochastic sensor characteristics, which can be obtained from factory or in-lab calibration through methods such as Allan variances \cite{El-Sheimy2008}. Therefore, setting of $\textbf{R}$ is more challenging because its elements are dependent on the quality of real-time measurements. Elements in $\textbf{R}$ may be set at empirical constants, by piece-wise models that have various but limited conditions, or by using adaptive Kalman filters that are based on variables such as residuals \cite{YuM2012}, innovations \cite{Aghili2016}, and the expectation-maximization term \cite{HuangY2018}. These methods are presented from the information fusion level. On the other hand, it is worthwhile to predict the accuracy of measurement models before information fusion, so as to obtain a priori information for the models in the filter.
        
 	For geometrical localization techniques (e.g., multilateration), criteria such as the dilution of precision (DOP) values are adopted to predict system accuracy, given device-AP distances and AP locations \cite{Langley1999}. There are also theoretical analyses for accuracy of angle-of-arrival \cite{HeY2015} and time-of-arrival/angle-of-arrival \cite{LiYGQ2018} based localization systems. 
 	
 	For fingerprinting, most existing works evaluate its accuracy through field testing (i.e., by localizing with real data and comparing with reference values) \cite{Yiu2016} or simulation \cite{Nguyen2017}. These methods can provide statistics of location errors at various points. However, collection of reference values for field testing is commonly time-consuming, labor-costly, and unaffordable for low-cost applications, while simulation cannot reflect real-world error sources. Researchers have also introduced theoretical methods, such as that uses the Cramer-Rao lower bound (CRLB) to estimate the lower bound for the accuracy of RSS databases \cite{Nikitin2017}, and that uses observability analysis \cite{Lourenco2013} to analyze RSS based localization. Both CRLB and observability analysis methods can reveal details inside the system theoretically. However, CRLB provides a lower bound instead a real-time prediction, while theoretical observability analysis can only analyze simplified models but cannot consider complex terms such as non-linear terms and noises. 
 	
 	For fingerprinting accuracy prediction using real-time data, the research \cite{Lemelson2009} proposes the initial work that predicts position errors through methods such as that uses the distances between one fingerprint and its closest fingerprints in database. Meanwhile, there are fingerprinting accuracy prediction approaches such as that uses DOP-like values \cite{Moghtadaiee2012}, Gaussian distribution based indicators \cite{Marcus2013}\cite{Beder2011}, and entropy from information theory \cite{Berkvens2018}. The fingerprinting accuracy prediction criteria in this paper follow the research \cite{Lemelson2009} but have extensions according to the crowdsourcing-based multi-sensor integration application, as discussed in the next subsection.
 	\begin{figure}
           \centering
           \includegraphics[width=0.45 \textwidth]{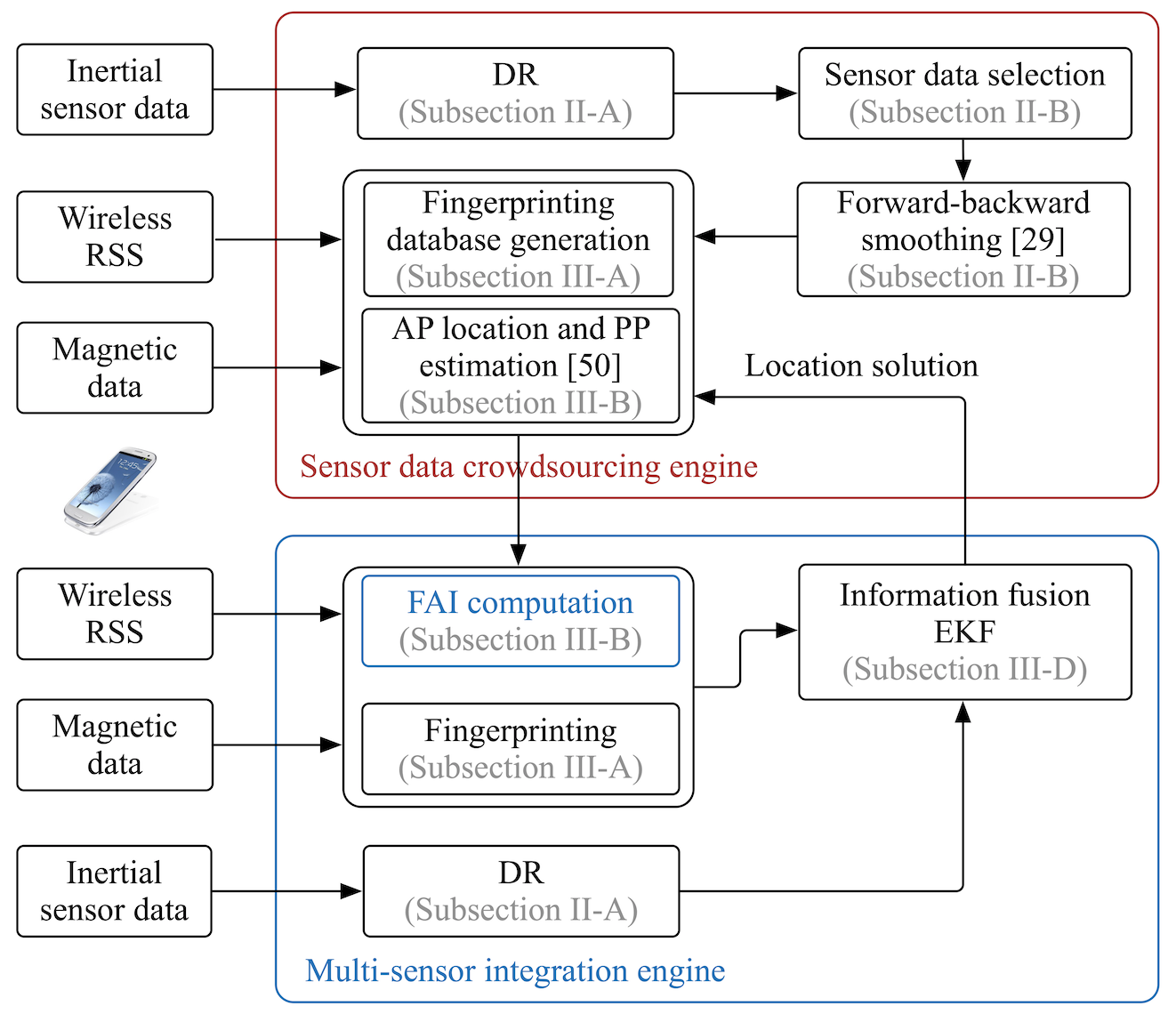}
           \caption{Diagram of proposed crowdsourcing-based localization method  }
           \label{fig:sys-diag}
         \end{figure} 
 	
         \subsection{Main Contributions}               
         Compared to the existing works, the main contributions of this paper include
         \begin{itemize}
         \item The requirement for costly database-training process has limited the use of fingerprinting (e.g., wireless and magnetic fingerprinting) methods in mass-market location-enhanced IoT applications. Thus, this paper proposes a crowdsourcing-based localization approach, which does not involve user intervention or parameter tuning. Crowdsourced sensor data is used for updating both wireless and magnetic databases simultaneously during the localization process.   
         \item It is not straightforward to set and tune the uncertainty of fingerprinting solutions in multi-sensor integration. Therefore, this paper introduces the fingerprinting accuracy indicator (FAI) factors from three levels (i.e., signal strength, geometry, and database levels) as well as their combinations. Especially, the weighted distance between similar fingerprints (weighted DSF) based FAI is proposed to provide robust predictions from the database level. Additionally, the FAI factors for magnetic fingerprinting are designed. Furthermore, the FAI-enhanced extended Kalman filter (EKF) is proposed, which is effective in enhancing both localization accuracy and reliability (i.e., the capability to resist location outliers). Based on the proposed crowdsourcing localization method, the advantages and limitations of each FAI factor on wireless and magnetic fingerprinting are investigated. It is found that the weighted DSF based FAI is effective in predicting both location errors and outliers. Meanwhile, the geometry based FAI can predict short-term errors and outliers but does not provide accurate tracking of long-term errors. In contrast, the signal strength based FAI is relatively stronger in predicting long-term location errors. Therefore, the presented FAI factors have complementary characteristics and thus are combined to provide more accurate and robust predictions.  
         \end{itemize}
         
    		\subsection{System Description}		
    		Figure \ref{fig:sys-diag} demonstrates the system diagram for the proposed multi-sensor crowdsourcing-based localization method. There are mainly two parts: the crowdsourcing engine and the FAI-enhanced multi-sensor integration engine, which are highlighted by the red and blue boxes, respectively. Both engines process raw sensor (e.g., inertial, wireless, and magnetic sensors) data from IoT devices. The crowdsourcing engine updates wireless and magnetic fingerprint databases with robust sensor data, which is selected through a quality assessment framework that is proposed in \cite{ZhangP2018} and furthermore improved in this paper. AP locations and PPs are also estimated and further used to calculate the geometry based FAI factor.
    		Within the multi-sensor integration engine, both wireless RSS and magnetic fingerprinting solutions are used as position updates, while the DR data is used to construct the system model. Data from fingerprinting and DR are fused in the FAI-enhanced EKF. Compared to traditional multi-sensor systems, the proposed EKF has an extra the FAI computation module, which predicts the accuracy (or uncertainty) of fingerprinting solutions in real time and thus removes the complex parameter-tuning process. 
    		
    		The details of the proposed system are described in the following sections. Section \ref{sec-sen-data-crowd} illustrates the sensor-based DR that is used for generation of crowdsourced RPs. Section \ref{fai-enh-ms-loc} describes the FAI-enhanced multi-sensor integration approach. Section \ref{sec:test} shows the tests and analyses, and Section \ref{sec-conclusionsl-work} draws the conclusions.
    		\begin{table}
           \centering
           \begin{tabular}{p{1.6cm} p{5.8cm} }
             \hline
             \textbf{Abbreviation} & \textbf{Definition}  \\ \hline
            AP  &  access point \\ 
			CDF  & cumulative distribution function \\ 
			CRLB & Cramer-Rao lower bound \\ 
			CT   & constant fingerprinting accuracy \\ 
			DR   & dead-reckoning \\ 
			DSF  & distance between similar fingerprints \\ 
			EKF  & extended Kalman filter  \\ 
			FAI  & fingerprinting accuracy indicators \\ 
			GPS  & global positioning system \\ 
			INS  & inertial navigation systems  \\ 
			IoT  & internet of things \\ 
			LPWAN  & low-power wide-area network \\ 
			﻿MC   & ﻿multi-FAI combination through linear combination \\ 
			MCM  & ﻿multi-FAI combination through selection of maximum value \\ 
			PDR  & pedestrian dead-reckoning \\ 
			PP   & path-loss model parameter \\ 
			RGB-D & red-green-blue-depth \\ 
			RMS   & root mean square \\ 
			RP    & reference points \\ 
			RSS   & received signal strength \\ 
			SD    & geometry based FAI \\ 
			SLAM  & simultaneous localization and mapping \\ 
			SS    & signal strength based FAI \\ 
			STD   & standard deviation \\ 
			UTC   & coordinated universal time \\ 
			WD    & weighted DSF based FAI, WDSF \\ 
			WiFi  & wireless local area network      \\ \hline
           \end{tabular}
           \caption{ List of abbreviations   }
           \label{tab:abbre}
         \end{table}
             		
         \section{Sensor Data Crowdsourcing}
         \label{sec-sen-data-crowd}	
         As described in Subsection \ref{sub-fp}, a fingerprinting database consists of a set of fingerprints (i.e., [feature vector, RP location] pairs). In this paper, wireless RSS and magnetic data are used to extract the features, while sensor-based DR generates the RP locations. Refer to \cite{ZhuangY2016} and \cite{LiY2017} for details about RSS and magnetic data preprocessing and feature extraction, respectively. This paper focuses on obtaining reliable RP locations in crowdsourcing. Compared to \cite{ZhangP2018}, this paper improves the condition for sensor data selection and avoids the complex model-training process. This section provides the methodology for sensor-based DR and sensor data selection.
         
          \subsection{Sensor-Based Dead-Reckoning}
          In contrast to the research \cite{LiY2017} that uses an attitude-determination EKF and a position-tracking EKF, this paper uses an inertial navigation systems (INS)/pedestrian dead-reckoning (PDR) integrated EKF for localization. The EKF system and measurement models are described in this subsection.
          
          \subsubsection{EKF System Models}
          The simplified motion model in \cite{Shin2005} is applied as the continuous system model as
	    	\begin{equation}
	    	\begin{aligned}
	   	\left[
  		\begin{matrix}
  		 \delta \dot{\textbf{p}}^n \\
  		 \delta \dot{\textbf{v}}^n \\
  		 \delta \dot{\bm{\psi}} \\
  		 \dot{\textbf{b}}_g \\
  		 \dot{\textbf{b}}_a
  		 \end{matrix} 
  		 \right]	 
  		 = 
  		 \left[
  		\begin{matrix}
  		-[\bm{\omega}^n_{en} \times] \delta \textbf{r}^n + \delta \textbf{v}^n \\
  		-[(2\bm{\omega}^n_{e} + \bm{\omega}^n_{en}) \times] \delta\textbf{v}^n + [\textbf{f}^n  \times] \bm{\psi} + \textbf{C}^n_b(\textbf{b}_a + \textbf{w}_a) \\
  		-[(\bm{\omega}^n_{e} + \bm{\omega}^n_{en}) \times ]\bm{\psi} - \textbf{C}^n_b(\textbf{b}_g + \textbf{w}_g) \\
  		- (\frac{1}{\bm{\tau}_{b_g}})\textbf{b}_g + \textbf{w}_{b_g} \\
  		- (\frac{1}{\bm{\tau}_{b_a}})\textbf{b}_a + \textbf{w}_{b_a}
  		\end{matrix} 
  		 \right]	 
  		 \end{aligned}
	    	\end{equation}
	    	where the states $\delta \textbf{p}^n$, $\delta \textbf{v}^n$, $\bm{\psi}$, $\textbf{b}_g$, and $\textbf{b}_a$ are the vectors of position errors, velocity errors, attitude errors, gyro biases, and accelerometer biases, respectively; $\textbf{C}^n_b$ is the direction cosine matrix from the device body frame (i.e., $b$-frame) to the local level frame (i.e., $n$-frame) predicted by the INS mechanization (refer to \cite{Shin2005} for the INS mechanization and the definition of frames); $\textbf{f}^n$ is the specific force vector projected to the $n$-frame, and $\bm{\omega}^n_{e}$ and $\bm{\omega}^n_{en}$ represent the angular rate of the Earth and that of the $n$-frame with respect to the Earth frame (i.e., e-frame), respectively; $\textbf{w}_g$ and $\textbf{w}_a$ are noises in gyro and accelerometer readings, respectively; $\bm{\tau}_{b_g}$ and $\bm{\tau}_{b_a}$ denote for the correlation time of sensor biases; and $\textbf{w}_{b_g}$ and $\textbf{w}_{b_a}$ are the driving noises for $\textbf{b}_g$ and $\textbf{b}_a$. The sign $[\bm{\gamma} \times]$ denotes the cross-product (skew-symmetric) form of the three-dimensional vector $\bm{\gamma}=\left[ \begin{matrix} \gamma_1 & \gamma_2 & \gamma_3 \end{matrix}  \right]^T$,
	    \begin{equation}
	    \begin{aligned}
  		&[\bm{\gamma} \times] = 
  		\left[
  		\begin{matrix}
  		 0 & -\gamma_3 & \gamma_2 \\
   		 \gamma_3 & 0 & -\gamma_1    \\
  		 -\gamma_2 & \gamma_1 & 0   \\
  		 \end{matrix} 
  		 \right]
  		 \end{aligned}
	    \end{equation}	 
			    	  
         \subsubsection{EKF Measurement Models}    
         Two types of updates are used in the EKF, including the sensor-related updates and motion-related updates. The sensor-related updates include the accelerometer and magnetometer measurement updates \cite{LiY2017}
         \begin{equation}
	      \hat{\textbf{f}}^n-\tilde{\textbf{f}}^n = -[\textbf{g}^n \times]\bm{\psi} + \textbf{C}^n_b \bm{\iota}_f
	      \end{equation}
	      \begin{equation}
	      \hat{\textbf{m}}^n-\tilde{\textbf{m}}^n  = -[\textbf{m}^n \times]\bm{\psi} + \textbf{C}^n_b \bm{\iota}_m
	      \end{equation} 
	      where $\textbf{g}^n$ and $\textbf{m}^n$ are the local gravity and magnetic intensity vectors, respectively. The signs $\tilde{\bm{\gamma}}$ and $\hat{\bm{\gamma}}$ represent the term $\bm{\gamma}$ that is measured and predicted by the system model, respectively. $\bm{\iota}_f$ and $\bm{\iota}_m$ are the accelerometer and magnetometer measurement noise vectors. 
			
  		  The motion-related updates include the pedestrian velocity model \cite{Syed2008} and the zero angular rate updates, which are described as
  		  \begin{equation}
	      \hat{\textbf{v}}-\tilde{\textbf{v}} = (\textbf{C}^n_b)^T  \delta\textbf{v}^n - (\textbf{C}^n_b)^T [\textbf{v}^n \times] \bm{\psi} + \bm{\iota}_v
	      \end{equation}	   
	      \begin{equation}
	      \delta \bm{\omega}^b_{ib} = \textbf{b}_g + \bm{\iota}_{\omega}
	      \end{equation}	
	      where $\textbf{v}^n$ is the velocity in the $n$-frame and $\bm{\omega}^b_{ib}$ is the gyro measurement vector; $\bm{\iota}_v$ and $\bm{\iota}_{\omega}$ are the velocity and angular rate measurement noise vectors. When the device is static, the velocity update becomes the zero velocity update, and the zero angular rate update is activated.
	      
	      The measured velocity vector is obtained from PDR by 
	      \begin{equation}
	      \tilde{\textbf{v}} = [\frac{s_k}{t_k - t_{k-1}} ~ 0 ~ 0]^T
	      \end{equation}
	      where $s_k$ is the step length between time epochs $t_{k-1}$ and $t_k$. Refer to \cite{ZhangH2015} and \cite{ShinSH2007} for details about step detection and step length estimation, respectively. 
%

		\subsection{Crowdsourced Sensor Data Selection }
		\label{sec-sdqs}
		From the big data perspective, only a small proportion of crowdsourced data is reliable enough for database updating. In \cite{ZhangP2018}, three sensor data quality assessment conditions are presented to select reliable sensor data: a) using the localization data that lasts shorter than a time threshold (e.g., 10 minutes); b) using the trajectory segments in which both start and end points are anchor points (i.e., points that have known coordinates provided by absolute localization technologies such as satellite positioning and Bluetooth); and c) using forward-backward smoothing to improve DR accuracy. In this paper, the conditions b) and c) are kept, while a) is changed to a) the position errors (compared to the location of anchor points) at the end points of both forward and backward DR solutions are within a geographical distance threshold $\lambda_d$. Through this change, the condition is changed from raw data to location results, and thus becomes stricter. The benefit for using the new condition is that it becomes straightforward to quantify the RP location uncertainty, which is further contained in the FAI values to improve the FAI prediction accuracy. 
		
		Additionally, although the research \cite{ZhangP2018} has proposed the sensor data quality assessment conditions, it has not involved the expected RP location uncertainty, which is essential for predicting the uncertainty for the generated fingerprinting database. Thus, this paper estimates the RP location uncertainty as follows. The equation that is used for computation of the forward-backward smoothed DR solution is
		\begin{equation}
		\begin{aligned}
		\textbf{x}_{sm,k} &= \frac{u_{\emph{+},k}}{u_{\emph{+},k}+u_{\emph{-},k}} \textbf{x}_{\emph{+},k} + \frac{u_{\emph{-},k}}{u_{\emph{+},k}+u_{\emph{-},k}} \textbf{x}_{\emph{-},k} \\
		              &=\xi_k \textbf{x}_{\emph{+},k} +(1-\xi_k) \textbf{x}_{\emph{-},k}, ~ \xi_k = \frac{u_{\emph{+},k}}{u_{\emph{+},k}+u_{\emph{-},k}}
		\end{aligned}
		\end{equation}
		where the subscripts $\emph{+}$, $\emph{-}$, and $sm$ indicate the forward, backward, and smoothed solutions; $\textbf{x}$ is the state vector, and $u$ is the weight; the subscript $k$ indicates the data epoch.
		
		According the error propagation law, the accuracy for the smoothed DR solution is 
		\begin{equation}
		\label{eq-sm-accu}
		\bm{\sigma}_{sm,k} = \sqrt{ \xi^2_k \bm{\sigma}^2_{\emph{+},k} + (1-\xi_k)^2 \bm{\sigma}^2_{\emph{-},k}}
		\end{equation}
		where $\bm{\sigma}$ represents the accuracy (i.e., the root mean square (RMS) value). Through Equation \eqref{eq-sm-accu}, it is possible to approximately predict the RP location uncertainties, which are used when computing FAI factors in the next section.		
%
		
		\section{FAI-Enhanced Multi-Sensor Integrated Localization}    
		\label{fai-enh-ms-loc}		    		
    		\subsection{Fingerprinting}	
    		\label{sec-fp1}
    		The Gaussian distribution based fingerprinting method \cite{Haeberlen2004} computes the probability density function of the states according to given measurements. The index for the selected fingerprint is computed by
		\begin{equation}
		\label{eq-i}
		\begin{aligned}
		\hat{i} & = \mathop{\arg\max}\limits_{i}  \left( {\zeta(\textbf{l}_i | \textbf{q})}  \right)
		 = \mathop{\arg\max}\limits_{i} \left(  {  \frac{ \zeta(\textbf{l}_i) \zeta(\textbf{q} | \textbf{l}_i)} {\zeta(\textbf{q})}}   \right)   \\
		& = \mathop{\arg\max}\limits_{i} \left( { \zeta(\textbf{l}_i) \zeta(\textbf{q} | \textbf{l}_i) }	   \right) \\
		& = \mathop{\arg\max}\limits_{i} \left(  { \zeta(\textbf{l}_i) {  \prod_{j=1}^{n} {\zeta(q_j|l_{i,j})}  } }   \right)
		\end{aligned}
		\end{equation}
		where $\zeta()$ represents the likelihood, $\textbf{q}$ denotes the measured feature vector, $\textbf{l}_i$ represents the reference feature vector at fingerprint $i$ in the database; $q_j$ and $l_{i,j}$ are the $j$-th feature in $\textbf{q}$ and $\textbf{l}_i$, respectively; $n$ is the dimension for the feature vector. The likelihood ${\zeta(q_j|l_{i,j})}$ can be calculated through the Gaussian distribution model as 
		\begin{equation}
		\zeta(q_j | { l_{i,j} } )= {  \frac{1}{\sigma_{i,j}\sqrt{2\pi}} \exp \left(    {-\frac{(q_{j}-\mu_{i,j})^2}{2\sigma_{i,j}^2}}  \right)  }
		\end{equation}		
		where $\mu_{i,j}$ and $\sigma_{i,j}^2$ are the mean and variance of feature $j$ in fingerprint $i$, respectively.
		   				   		
    	For RSS and magnetic fingerprinting, the features are RSS from multiple APs and the horizontal and vertical magnetic gradient profiles, respectively. The method to generate the magnetic gradient profile is as follows. First, the magnetic intensity profile is built by combining magnetic intensity values at a series of points on a short-term trajectory (e.g., 10 steps). To increase the magnetic fingerprint dimension, accelerometer-derived horizontal angles are used to extract the horizontal and vertical magnetic components. Afterwards, each element in the magnetic intensity profile is subtracted by its first element to generate the magnetic gradient profile. Refer to \cite{LiY2017} for details about RSS and magnetic fingerprints. The magnetic matching solutions are obtained through the wireless-aided magnetic matching strategy from \cite{LiY2017}. That is, a circle area is first determined by wireless localization. The circle center locates at the wireless location solution, while the circle radius is set at three times of wireless location accuracy. Then, magnetic fingerprinting is implemented within the circle. 
    	
    	Once the likelihood for each RP is calculated, $\kappa$ RPs that  have the highest likelihood values are selected. Afterwards, the location of selected RPs are weighted averaged to calculate the position estimate by 	
		\begin{equation} 
		\label{eq-knn}
		\hat{\textbf{p}} = \sum_{i=1}^{\kappa} {{ \frac{\textbf{p}_i  \zeta_i }{ \sum_{j=1}^{\kappa} {\zeta_j} }} }
		\end{equation}
		 where $\hat{\textbf{p}}$ is the position estimate, $\textbf{p}_i$ is the location for the $i$-th RP and $\zeta_i$ is its likelihood.  		
    		
		 \subsection{Fingerprinting Accuracy Indicator}
		 FAI at three levels are introduced, including the signal strength, geometry, and database DSF levels. This subsection describes these FAI factors and their combinations.

	     \subsubsection{Signal Strength Based FAI (SS)}
	     \label{sec-ss}
	     Accordingly to preliminary tests, features with stronger RSS or magnetic gradient values are generally more reliable. Specifically, a stronger RSS indicates a closer distance to an AP, and stronger magnetic gradients indicate the fingerprint has a higher probability to be indistinct from others. Based on this principle, each feature is given a score according to its value. Afterwards, scores from all features are combined to determine the SS value by
	     \begin{equation}
	     \label{eq-ss-alpha}
           [ss]^* = \alpha^*_{ss}\frac{\sum_{i=1}^{n_*} {c^*_{i}}}{n_*} 
         \end{equation}    
         where the superscript and subscript $*$ may be $w$ for wireless RSS or $m$ for the magnetic data; $n_*$ is the number of features; $c^*_{i}$ is the score for feature $i$, and $\alpha^*_{ss}$ is the scale factor for SS computation. The $c^*_{i}$ values for RSS and magnetic data are respectively computed as
         \begin{equation}
           c^w_{i} = \frac{1}{d_{i}} = 10^{ \frac{r_i-\beta_{2,i}}{10\beta_{1,i}}  }, ~ c^m_{i} = \frac{1}{m_i}
         \end{equation}
%
         where $d_i$ is the distance from device to AP $i$, $r_i$ represents the RSS from AP $i$; $\beta_{1,i}$ and $\beta_{2,i}$ are PPs for AP $i$, and $m_i$ represents the value of magnetic gradient feature $i$. The PPs $\beta_{1}$ and $\beta_{2}$ for each AP are determined through least squares by  using the measurement model in \cite{ZhuangWcl} as
         \begin{equation} \label{rss-hx}
         \begin{aligned}
         \textbf{r}_a = -10 \beta_1 log_{10}{\left(  \sqrt{(x-\textbf{x}_u)^2 + (y-\textbf{y}_u)^2}  \right)} + \beta_2
         \end{aligned}
         \end{equation}
         where $\textbf{x}_u=\left[ \begin{matrix}  x_{u,1} & x_{u,2} & ... & x_{u,n_p}  \end{matrix}  \right]^T$ and $\textbf{y}_u=\left[ \begin{matrix}  y_{u,1} & y_{u,2} & ... & y_{u,n_p}  \end{matrix}  \right]^T$ are device coordinates from DR; $\textbf{r}_{a} = \left[ \begin{matrix}  r_1 & r_2 & ... & r_{n_p}  \end{matrix}  \right]^T$ is the RSS measurement vector, and $n_p$ is the number of device locations. The state vector to be estimated is $\textbf{x}_{a} =\left[ x~~y~~\beta_1~~\beta_2 \right]^T$, where $x$ and $y$ are the AP coordinates and $\beta_1$ and $\beta_2$ are the AP PPs. The design matrix $\textbf{H}_a$ for AP location and PP estimation is  
	    \begin{equation}
		\begin{aligned}
  		&\textbf{H}_{a} = 
  		\left[
  		\begin{matrix}
  		 \frac{-10n(x-x_{u,1})}{d^2_1 ln10} & \frac{-10n(y-y_{u,1})}{d^2_1 ln10} & -10log_{10}{d_1} & 1 \\
   		 ... & ... & ... & ...   \\
  		 \frac{-10n(x-x_{u,j})}{d^2_j ln10} & \frac{-10n(y-y_{u,j})}{d^2_j ln10} & -10log_{10}{d_j} & 1 \\
  		 ... & ... & ... & ...   \\
  		 \frac{-10n(x-x_{u,n_p})}{d^2_{n_p} ln10} & \frac{-10n(y-y_{u,n_p})}{d^2_{n_p} ln10} & -10log_{10}{d_{n_p}} & 1 \\
  		 \end{matrix} 
  		 \right]
  		 \end{aligned}
         \end{equation}
                
		Using least squares, the state vector $\textbf{x}_a$ and corresponding covariance matrix $\textbf{P}_{a}$ are estimated by
		\begin{equation} 
		\label{eq-ls}
          \hat{\textbf{x}}_{a} = \left( \textbf{H}^T_{a} \textbf{R}^{-1}_{a} \textbf{H}_{a}  \right)^{-1}\textbf{H}^T_{a}  \textbf{R}^{-1}_{a} \textbf{r}_{a}
         \end{equation}
         \begin{equation} 
		\label{eq-ls-p}
          \hat{\textbf{P}}_{a} = \left( \textbf{H}^T_{a} \textbf{R}^{-1}_{a} \textbf{H}_{a}  \right)^{-1} 
         \end{equation}
         where $\textbf{R}_{a} =diag(\bm{\iota}^2_{r} )$ and $\bm{\iota}_{r} $ is the RSS noise vector. The sign $diag(\bm{\gamma})$ represents the diagonal matrix in which the diagonal elements are the elements in $\bm{\gamma}$.

	     \subsubsection{Geometry Based FAI (SD)} 
	     For multilateration, localization accuracy is correlated with the measurement geometry that can be quantified by the DOP value \cite{Langley1999}. Thus, the scaled DOP value is used as a FAI factor. Least squares is used to compute the DOP value. The design matrix for range-base DOP calculation is
         \begin{equation}
		\begin{aligned}
  		&\textbf{H}^w_{d} = 
  		\left[
  		\begin{matrix}
  		 \frac{x_u-x_{w,1}}{d_1} & \frac{y_u-y_{w,1}}{d_1} \\
   		 ... & ...   \\
   		 \frac{x_u-x_{w,j}}{d_j} & \frac{y_u-y_{w,j}}{d_j} \\
  		 ... & ...    \\
  		 \frac{x_u-x_{w,{n_w}}}{d_{n_w}} & \frac{y_u-y_{w,{n_w}}}{d_{n_w}}
  		 \end{matrix} 
  		 \right]
  		 \end{aligned}
         \end{equation}
         where $[x_u, y_u]$ and $[x_{w,j},y_{w,j}]$ are the locations for the device and AP $j$, respectively; $d_j$ is the distance from device to AP $j$, and $n_{w}$ is the dimension of the RSS feature vector. 
         
         A similar geometrical indicator is presented for magnetic gradient fingerprints. The design matrix is
		 \begin{equation}
		\begin{aligned}
  		&\textbf{H}^m_{d} = 
  		\left[
  		\begin{matrix}
  		 \frac{m_{h,1}}{\sqrt{m^2_{h,1} + m^2_{v,1}}} & \frac{m_{v,1}}{\sqrt{m^2_{h,1} + m^2_{v,1}}} \\
   		 ... & ...   \\
   		 \frac{m_{h,j}}{\sqrt{m^2_{h,j} + m^2_{y,j}}} & \frac{m_{v,j}}{\sqrt{m^2_{h,j} + m^2_{v,j}}} \\
  		 ... & ...    \\
  		 \frac{m_{h,n_m}}{\sqrt{m^2_{h,n_m} + m^2_{v,n_m}}} & \frac{m_{v,n_m}}{\sqrt{m^2_{h,n_m} + m^2_{v,n_m}}} \\
  		 \end{matrix} 
  		 \right]
  		 \end{aligned}
         \end{equation}
         where $m_{h,j}$ and $m_{v,j}$ are the $j$-th horizontal and vertical magnetic components, respectively; $n_{m}$ is the dimension of magnetic feature vector. 
         
         The matrix for DOP calculation is computed as
         \begin{equation} 
          \textbf{M}^*_{d} = \left( \textbf{H}^{*T}_{d} \textbf{H}^*_{d}  \right)^{-1} 
         \end{equation}
         
         To predict the localization accuracy, the scaled DOP value is calculated by 
         \begin{equation} 
         \label{eq-sd-alpha}
         [sd]^* = \alpha^*_{sd} \sqrt{{\textbf{M}}^*_{d}(1,1) + {\textbf{M}}^*_{d}(2,2)}
         \end{equation}
        where ${\textbf{M}}^*_{d}(i,i)$ is the $i$-th diagonal element in $\textbf{M}^*_{d}$ and $\alpha^*_{sd}$ is the scale factor for SD computation. 

	     \subsubsection{Weighted DSF Based FAI (WD)}
	     \label{sec-wd-fai}
		 According to \cite{Lemelson2009}, the accuracy of one fingerprint can be described by its DSF, that is, the mean value of geographical distances between this fingerprint and fingerprints that has similar features. Compared to \cite{Lemelson2009}, there are two differences in the FAI computation in this paper. First, the DSF is calculated by likelihood, instead of Euclidean distance. Second, the weighted DSF (WDSF, abbreviated as WD), instead of DSF, is used as the FAI. The equations for DSF computation are	
		\begin{equation}
		\label{eq-i}
		\begin{aligned}
		&[dsf]_{i} = \frac{1}{\kappa_d} { \sum_{k=1}^{\kappa_d} { {  \prod_{j=1}^{n} {  \left(  {  \frac{1}{\sigma_{i,j}\sqrt{2\pi}} \exp \left(    {-\frac{(l_{i_k,j}-\mu_{i,j})^2}{2\sigma_{i,j}^2}}  \right)  }    \right)   } } }   } 
		\end{aligned}
		\end{equation}
		where $[dsf]_{i}$ denotes the DSF value for fingerprint $i$; $i_k$ is the fingerprint that has the $k$-th closest features to fingerprint $i$; $\mu_{i,j}$ and $\sigma_{i,j}^2$ are the mean and variance of feature $j$ in fingerprint $i$; $\kappa_d$ is the number of similar fingerprints, and $n$ is the dimension of feature vector in fingerprints. 
		     
	     As shown by the results in Subsections \ref{sec-res-fai-wifi} and \ref{sec-res-fai-mag}, the DSF value may suffer from blunders in fingerprinting accuracy prediction. To improve the reliability, the WD of $\kappa$ selected fingerprints in Equation \eqref{eq-knn} are calculated by
	     \begin{equation}
	     [wd] = \frac{\sum_{i=1}^{\kappa} {\zeta_{i} [dsf]_{i}} } {\sum_{i=1}^{\kappa} {\zeta_{i}}}
	    \end{equation}
	    where $\zeta_i$ is the likelihood, which is calculated by the similarity between the measured feature vector and that in the $i$-th selected fingerprint in database.  
	    
	     As described in subsection \ref{sec-sdqs}, the crowdsourced RP location uncertainty $\bm{\sigma}_{sm}$ is added into each FAI by
          \begin{equation}
          \vartheta = \sqrt{\vartheta^2 + \bm{\sigma}^2_{sm}}, ~  \vartheta \in [[ss], [sd], [wd]]
         \end{equation}	
	    	   
	     \subsubsection{Multi-FAI Combination}
	     \label{sec-mfc}
	     SS, SD, and WD are proposed from different levels. Results in Subsections \ref{sec-res-fai-wifi} and \ref{sec-res-fai-mag} indicate that each factor has its own advantages. Thus, these factors may be combined to take better advantage of their merits. Two combination strategies are used. The first is denoted as MC, which is the linear combination as
	     \begin{equation}
	     \label{eq-lin-model-mc}
	     [mc]_k = \rho_{ss} [ss]_k + \rho_{sd} [sd]_k +  \rho_{wd} [wd]_k
	     \end{equation}  
	     where the subscript $k$ represents the data epoch; $\rho_{\vartheta}$ ($\vartheta \in [[ss], [sd], [wd]]$) are the coefficients for combination, which can be estimated through least squares by using DR solutions selected from the method in Subsection \ref{sec-sdqs}. 
	     
	     The second combination strategy is denoted as MCM, which is the maximum one among these values, that is   
	     \begin{equation}
	     [mcm]_k = max([ss]_k, [sd]_k, [wd]_k)
	     \end{equation}
	     
	     The maximum value is chosen because increasing the setting of measurement noises to a certain extent is a common approach in engineering practices. The reason for this fact is that practical measurement noises may contain systematic errors, which breaks the Gaussian noise assumption in EKF. In this case, using a larger measurement noise setting may enhance the capability to avoid degradation on EKF solutions.

	     \subsection{Multi-Sensor Integration with FAI}
	    As mentioned above, position solutions from fingerprinting are used as EKF updates. The measurement model is
	    \begin{equation}
	    \hat{\textbf{p}}^n-\tilde{\textbf{p}}^n = \delta\textbf{p}^n + \bm{\iota}_p
	    \end{equation}	    	 
	    	 where $\hat{\textbf{p}}$ and $\tilde{\textbf{p}}$ denotes the positions predicted by the system model and that obtained from fingerprinting; $\bm{\iota}_p$ is the measurement noise vector. The variance values of position measurement noises are embedded in the position measurement noise matrix $\textbf{R}_p$. The elements in the $\textbf{R}_p$ matrix are commonly set at empirical constants or by a piece-wise model, that is
	    	 \begin{equation}
 		 \textbf{R}_{p,k} = diag([\eta^2_{k}~~\eta^2_{k}]), 
 		 \end{equation}   
 		 \begin{equation}
 		 \label{eq-cnd1}
  		  \eta_{k}=\left\{
                \begin{array}{ll}
                  \eta_1~\ni~ cond.~ 1\\
                  \eta_2~\ni~ cond.~ 2\\
                  ... \\
                  \eta_i~\ni ~ cond.~ i\\
                \end{array}
              \right.
 		 \end{equation} 
 		 where the sign $cond.$ represents the conditions. Since there are limited number of conditions in Equation \eqref{eq-cnd1}, the measurement noises are constants in the short term. When there is an outlier (i.e., an instantaneous and significant location error) in the fingerprinting position update, the integrated system still sets a constant weight on the measurement. In this case, the integrated solution will be degraded. 
 		 
 		 To alleviate this issue, the FAI values are used to set the measurement noise as
 		  \begin{equation} 
          \eta_{k} = \vartheta_k, ~\vartheta \in [[ss], [sd], [wd], [mc], [mcm]]
         \end{equation}
 		 
 		 The details for EKF are described in the next subsection.

 		 \subsection{EKF Computation}
		The discrete-time EKF system and measurement models can be written in a general form as  
 		\begin{equation}
 		\textbf{x}_k = \varrho_{k,k-1}(\textbf{x}_{k-1}) + \textbf{w}_{k-1}, ~ \textbf{w}_{k-1} \sim N(\textbf{0}, \textbf{Q}_{k-1})
 		\end{equation}
 		\begin{equation}
 		\textbf{z}_k = h_{k}(\textbf{x}_{k}) + \bm{\iota}_k, ~ \bm{\iota}_k \sim N(\textbf{0}, \textbf{R}_{k})
 		\end{equation}
	where the subscripts $k$ and $k-1$ denote the data epochs; $\varrho()$ and $h()$ represents the system and measurement models, respectively; $\textbf{x}$ and $\textbf{z}$ are the state and measurement vectors; $\textbf{w}$ and $\bm{\iota}$ represent system and measurement noises, which have $\textbf{E}[\textbf{w}_i \textbf{n}_j]=0$ for all $i$ and $j$, where $\textbf{E}[~]$ denotes the expectation;  $\textbf{Q}$ and $\textbf{R}$ are the system and measurement noise covariance matrices, and the sign $N(\bm{\gamma}_1,\bm{\gamma}_2)$ denotes the Gaussian distribution that has mean of $\bm{\gamma}_1$ and variance of $\bm{\gamma}_2$.

	As the EKF estimates the process state and then obtains feedback from noisy measurements, it can be divided into prediction and update. In prediction, the EKF estimates the current states and uncertainties by
\begin{equation}
\textbf{x}^-_k= \bm{\Phi}_{k,k-1} \textbf{x}^+_{k-1} , ~ \bm{\Phi}_{k,k-1} \approx \frac{\partial \varrho_{k,k-1} }{\partial \textbf{x}_{k-1} }
\end{equation}  
\begin{equation}
\textbf{P}^-_k= \bm{\Phi}_{k,k-1} \textbf{P}^+_{k-1} \bm{\Phi}^T_{k,k-1} + \textbf{Q}_{k-1}
\end{equation}
	where $\bm{\Phi}$ is the state transition matrix and $\textbf{P}$ is the state error covariance matrix; the superscripts $-$ and $+$ indicates predicted and updated terms, respectively. Once a measurement is observed, the estimates are updated through the introduction of the Kalman filter gain $\textbf{K}_k$ by
\begin{equation}
\label{eq:kf-gain}
\textbf{K}_k= \textbf{P}^-_k \textbf{H}^T_k ( \textbf{H}_k \textbf{P}^-_k \textbf{H}^T_k + \textbf{R}_k    ) ^{-1}, ~ \textbf{H}_k \approx \frac{\partial h_{k} }{\partial \textbf{x}_{k} }
\end{equation} 
\begin{equation}
\label{eq:kf-gain2}
\textbf{x}^+_k=  \textbf{x}^-_k + \textbf{K}_k (\textbf{z}_{k} - \textbf{H}_k \textbf{x}^-_k)
\end{equation} 
\begin{equation}
\textbf{P}^+_k=  (\textbf{I} - \textbf{K}_k \textbf{H}_k ) \textbf{P}^-_k
\end{equation} 
	where $\textbf{H}$ is the design matrix and $\textbf{I}$ is an identify matrix. From Equations \eqref{eq:kf-gain} and \eqref{eq:kf-gain2}, EKF solutions are the weighted average of predictions and measurements. In the proposed method, the errors in fingerprinting position measurements are predicted by the FAI factors. Thus, if the FAI factors can provide a reliable prediction of the fingerprinting accuracy, the EKF will be intelligent to increase the elements in $\textbf{R}_k$ adaptively when there is an outlier. Accordingly, the values of elements in $\textbf{K}_k$ will be decreased and the impact of measurement errors will be mitigated.
 		
         \section{Tests and Results}
         \label{sec:test}
\begin{figure}
           \centering
           \includegraphics[width=0.48 \textwidth]{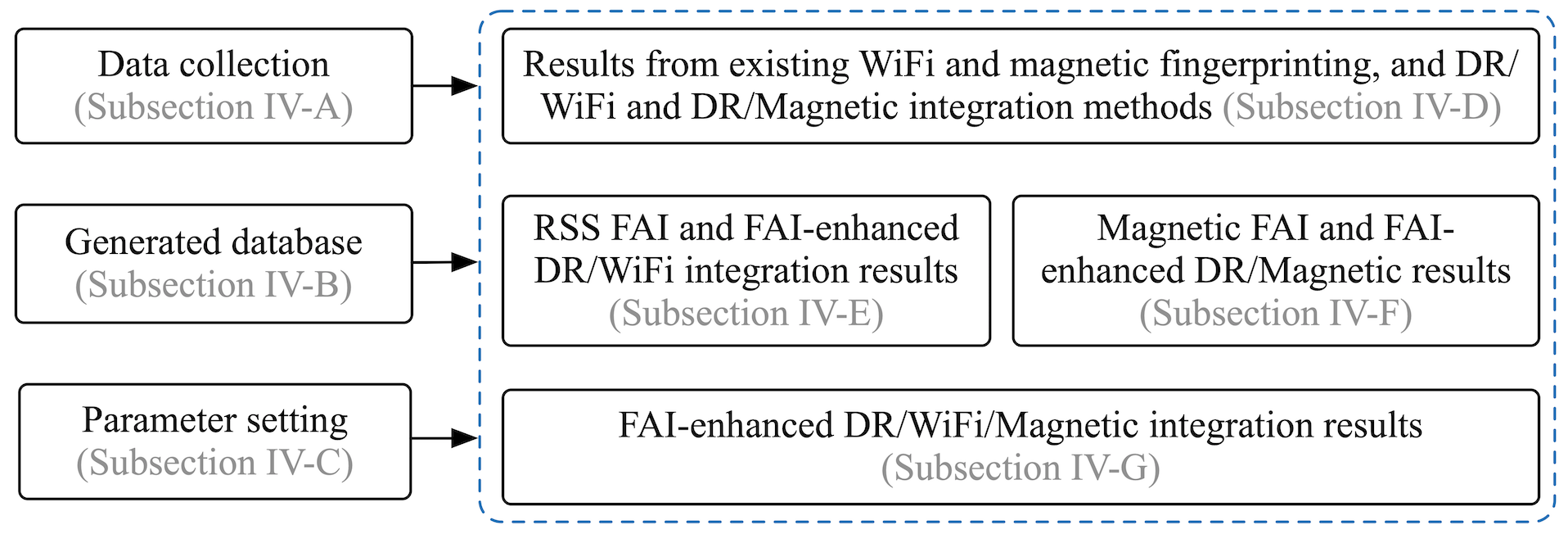}
           \caption{ Flow chart of tests  }
           \label{fig:flow-chart-test}
         \end{figure}

         Figure \ref{fig:flow-chart-test} shows the organization of this section. The test description is first provided, followed by the generated fingerprinting databases. Meanwhile, the parameter setting for the tests is illustrated. Afterwards, the results for WiFi and magnetic fingerprinting, the existing DR/WiFi, DR/Magnetic, and DR/WiFi/Magnetic integration methods, RSS and magnetic FAI prediction, and the FAI-enhanced methods are demonstrated.
         
         \subsection{Test Description}
\begin{figure}
           \centering
           \includegraphics[width=0.45 \textwidth]{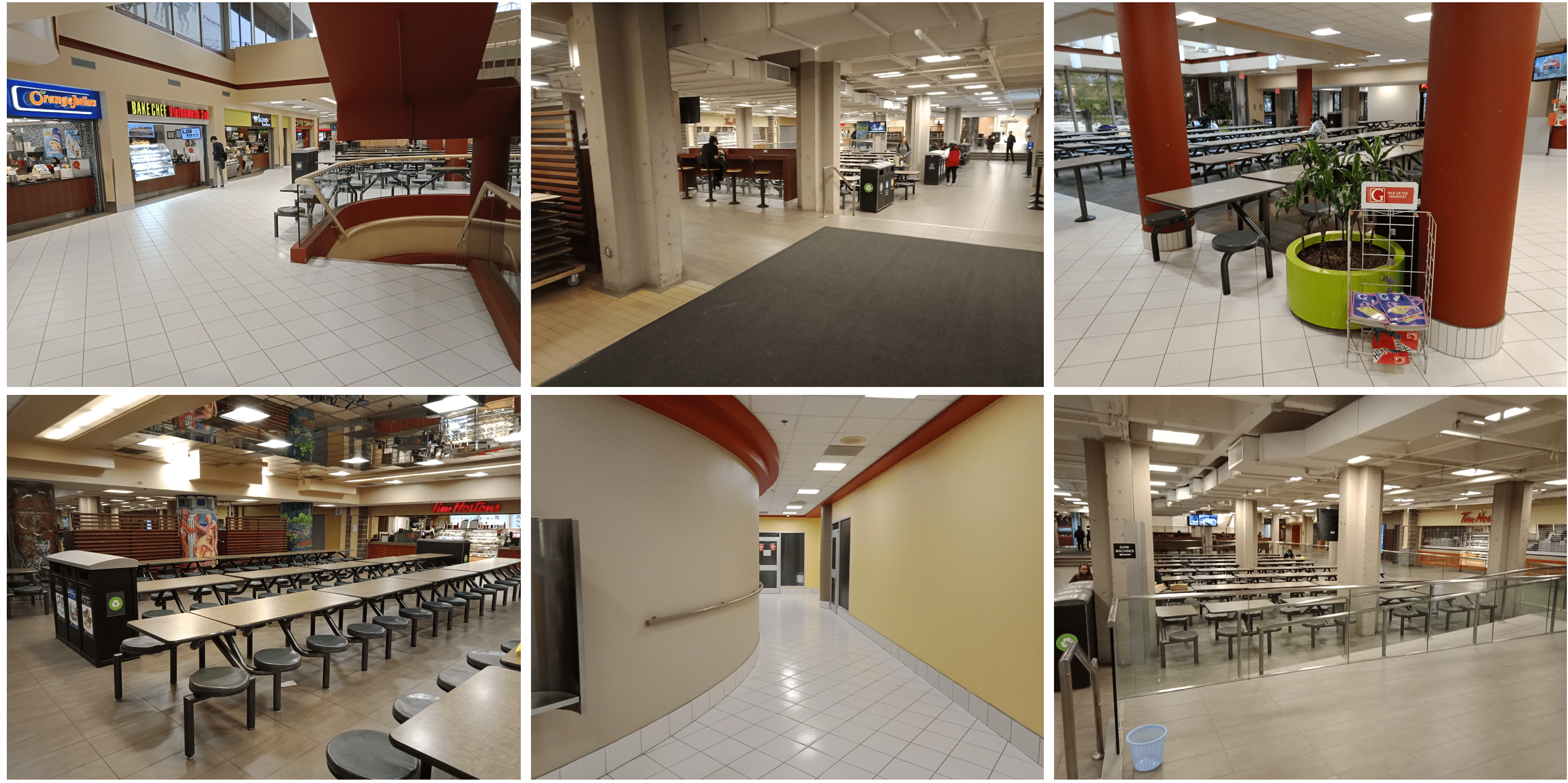}
           \caption{Test environment}
           \label{fig:machall}
         \end{figure}

         Field tests were conducted on the main floor of the MacEwan Student Center at the University of Calgary. The test area was a typical indoor shopping mall environment. Figure \ref{fig:machall} demonstrates pictures in the test area. There were stores around the space, and seats, walls, and open spaces in the middle area. Therefore, it was a complex indoor environment that involved both line-of-sight and non-line-of-sight areas, as well as open areas and corridors. Second, there were distinct magnetic features in some areas, and flat magnetic field in other areas. The size of test area was around 160 m by 90 m. 
         
         Five Android smart devices were used, including a Samsung Galaxy S4, a Galaxy S7, a Huawei P10, a Lenovo Phab 2 Pro phone, and a Nexus 9 tablet. All devices were equipped with three-axis gyros, accelerometers, and magnetometers, a WiFi receiver, and a global positioning system (GPS) receiver. The data rates were set at 20 Hz for for gyros, accelerometers, and magnetometers, 0.5 Hz for WiFi, and 0.1 Hz for GPS. All sensor data were logged into files for post-processing. Data from various sensors were synchronized by the coordinated universal time (UTC) time in Android.
                  
          For testing, five testers held the devices and each walked for four trajectories, each lasted for over 15 minutes. The reference trajectories were generated by taking a Lenovo Phab 2 Pro smartphone to collect red-green-blue-depth (RGB-D) images and generate SLAM solutions. The location accuracy of SLAM solutions was evaluated by using landmarks points from a commercial floor map in the test area. The SLAM location solutions had accuracy of 0.1 m and thus can be used as the references for the proposed method.

		\subsection{Generated Database}
\begin{figure}
           \centering
           \includegraphics[width=0.40 \textwidth]{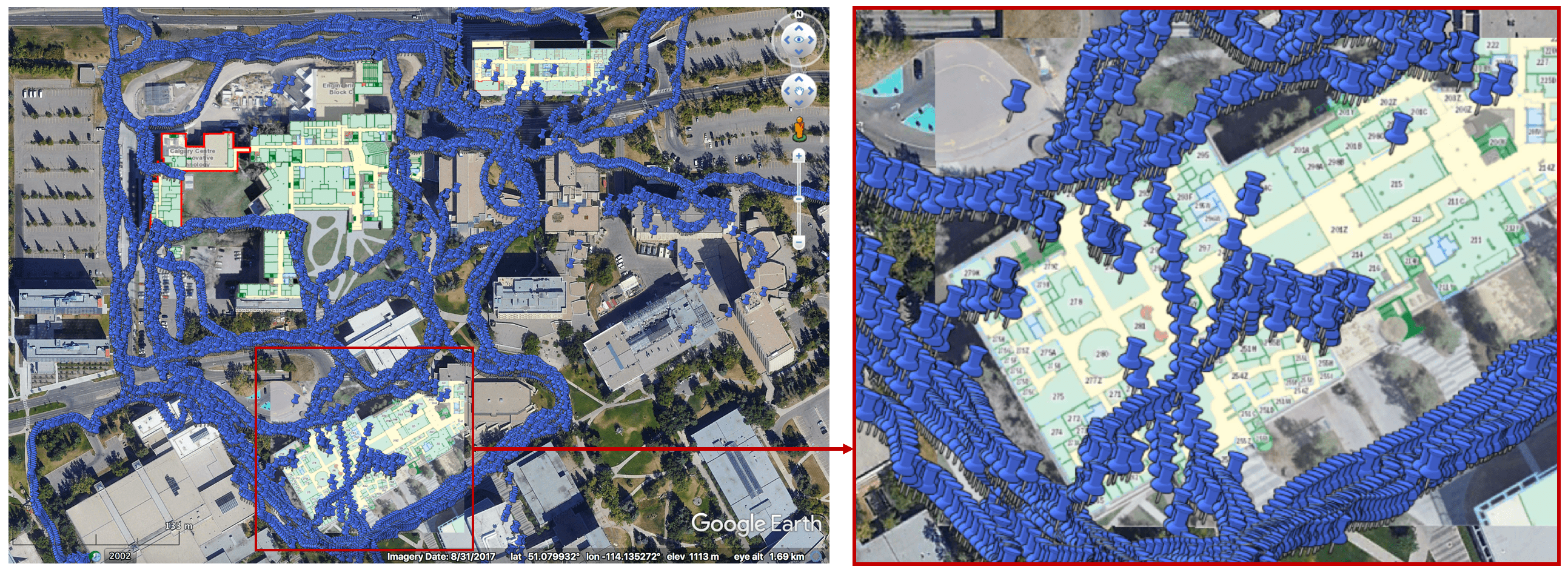}
           \caption{GPS trajectories with smartphones}
           \label{fig:gps}
         \end{figure}

         Five testers held the devices and walked randomly or guided around the test area for half an hour. Figure \ref{fig:gps} (left) shows the GPS localization points, while Figure \ref{fig:gps} (right) illustrates the zoomed-in figure, which contains the indoor map for the test area. GPS provided continuous localization solutions in outdoor areas, but suffered from signal outages indoors. Thus, it is difficult to use GPS to generate indoor RP locations. Sensor-based DR solutions can be used to fill this gap. 
         
\begin{figure}
           \centering
           \includegraphics[width=0.28 \textwidth]{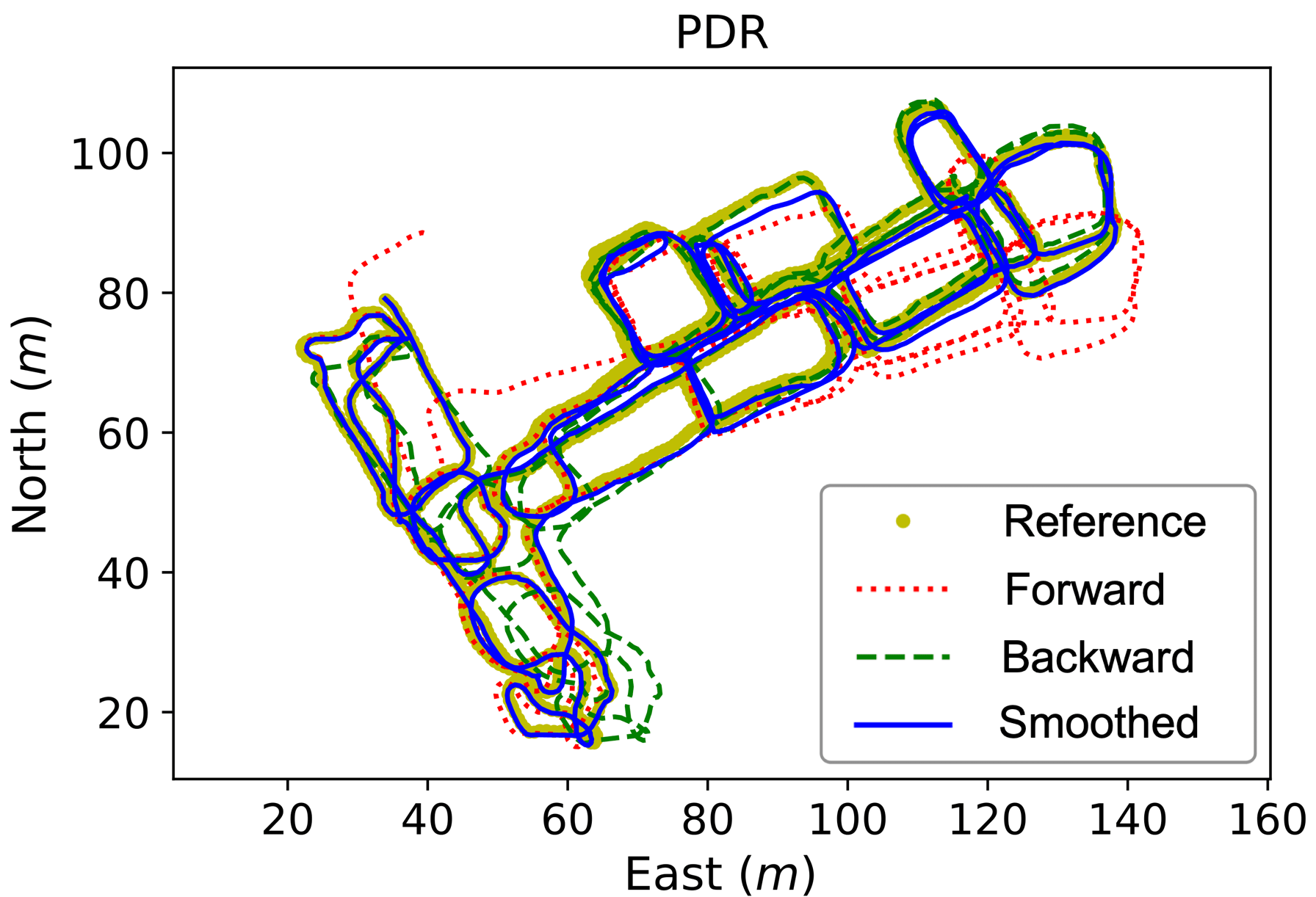}
           \caption{Example of smoothed DR trajectory}
           \label{fig:pdr}
         \end{figure}

         The crowdsourced sensor data selection approach in Subsection \ref{sec-sdqs} was used to select reliable sensor data. Figure \ref{fig:pdr} illustrates an example of the selected sensor data. Both start and end points of this path had GPS locations outdoors. If using the quality assessment conditions in \cite{ZhangP2018}, this data was not qualified because it lasted for a time period that was longer than the threshold (i.e., 10 minutes). However, this data was qualified under the new quality assessment condition because the localization errors (compared to GPS solutions) at the end of both forward and backward DR solutions were within the distance threshold $\lambda_d$ (i.e., 20 m).
         
           The smoothed DR locations were combined with the RSS and magnetic data to generate the fingerprinting databases. For this purpose, the two-dimensional space was first divided into grids that had a length of $l_g$ (i.e., 3 m). Then, fingerprints that located within each grid were combined to train the model for this grid. Both the mean and variance values for each feature were calculated. If the number of data within a grid was less than the threshold $\lambda_{n,1}$ (i.e., 5), the grid was discarded. Meanwhile, if the number of data within a grid was less than the threshold $\lambda_{n,2}$ (i.e., 20), a default variance value was used, which was (5 dBm)$^2$ for RSS and (0.03 Guass)$^2$ for the magnetic data.

\begin{figure}
           \centering
           \includegraphics[width=0.42 \textwidth]{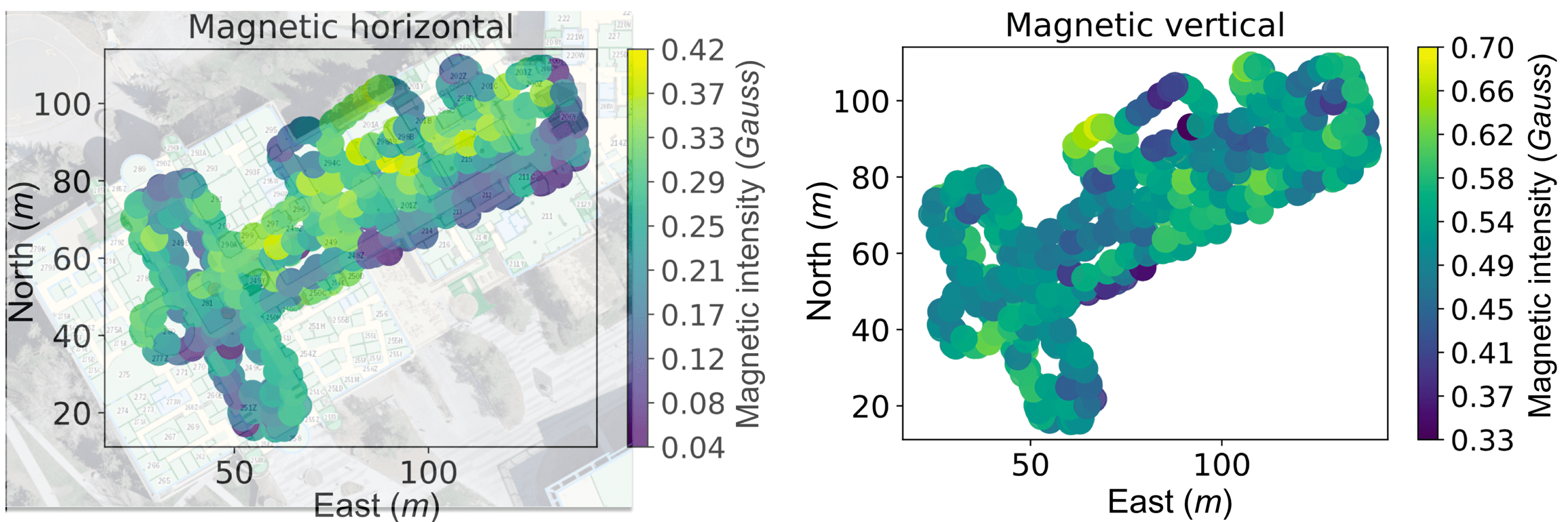}
           \caption{Magnetic map generated using crowdsourced data}
           \label{fig:mag-db}
         \end{figure}

\begin{figure}
           \centering
           \includegraphics[width=0.42 \textwidth]{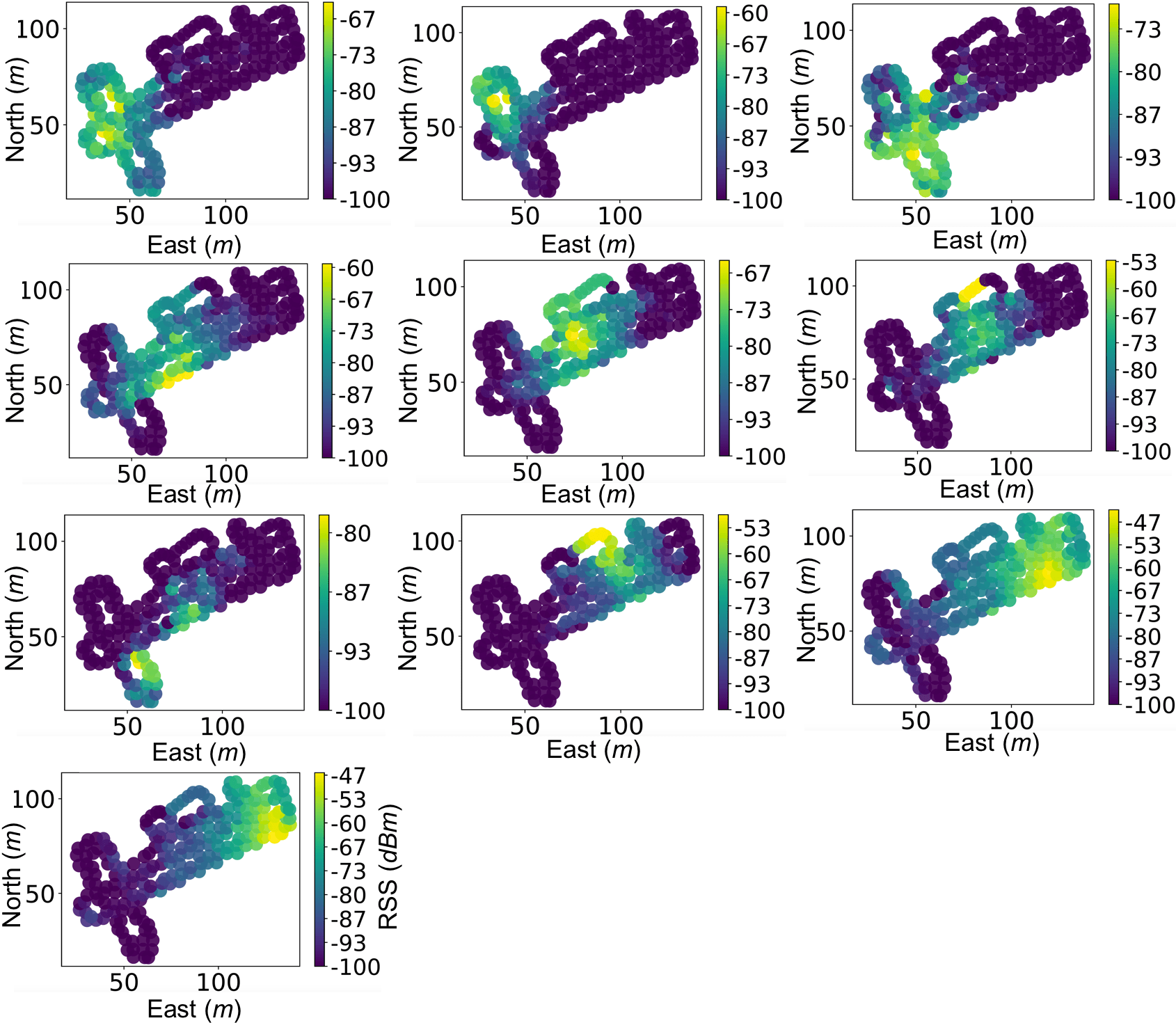}
           \caption{WiFi RSS map generated using crowdsourced data}
           \label{fig:wifi-db}
         \end{figure}

\begin{figure}
           \centering
           \includegraphics[width=0.45 \textwidth]{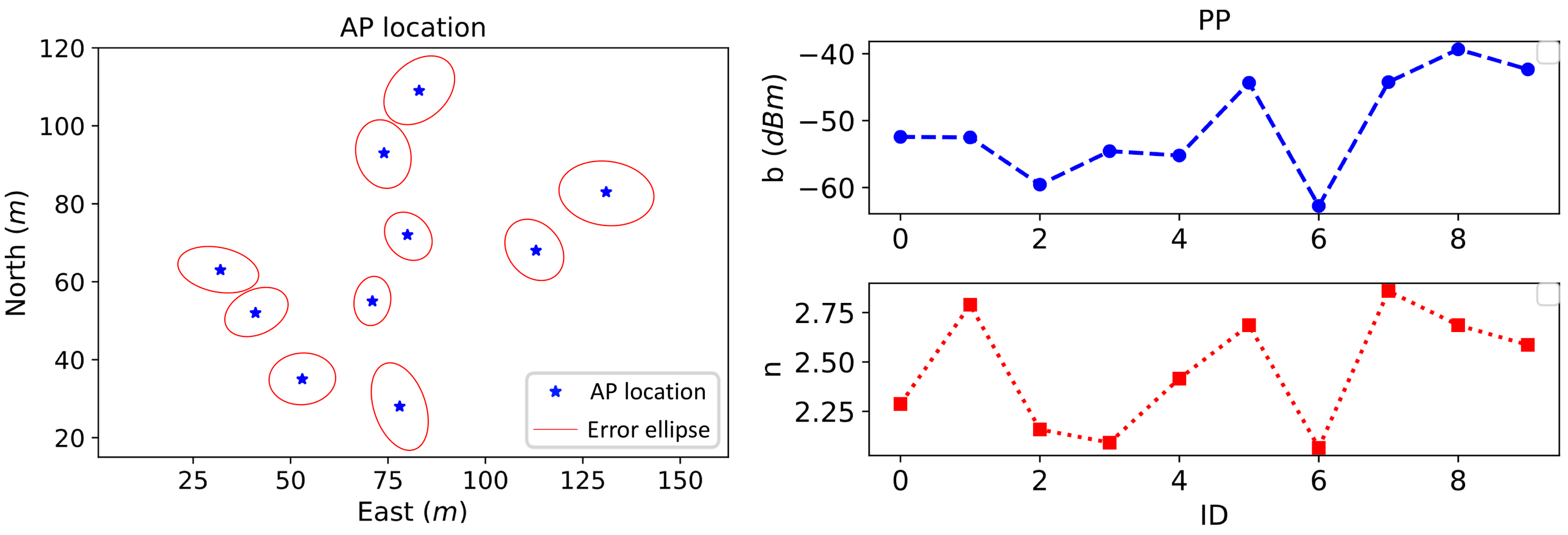}
           \caption{WiFi AP location and PP estimation solution}
           \label{fig:ap-pp}
         \end{figure}

		Figure \ref{fig:mag-db} demonstrates the distribution of the mean horizontal and vertical magnetic intensities over the space, and Figure \ref{fig:wifi-db} shows the signal distribution of 10 APs selected for localization. Figure \ref{fig:ap-pp} shows the estimated AP location (left) and PP values (right) through the method in subsection \ref{sec-ss}. The uncertainties for the estimated AP locations are demonstrated by ellipses. 
          
          \subsection{Parameter Setting}
          \label{sec-test-para}
		 The EKF parameters include the initial states (i.e., the initial position vector $\textbf{p}^n(0)$, velocity vector $\textbf{v}^n(0)$, attitude vector $\bm{\psi}(0)$, gyro bias vector $\textbf{b}_{g}(0)$, and  accelerometer bias vector $\textbf{b}_{a}(0)$), the initial EKF state vector (i.e., $\textbf{x}(0)$=$[\delta \textbf{p}^n(0) ~ \delta \textbf{v}^n(0)~ \delta \bm{\psi}(0)~ \textbf{b}_g(0)~  \textbf{b}_a(0)]$) and the corresponding initial state error covariance matrix (i.e., $\textbf{P}(0)$), the covariance matrix of system noises (i.e., $\textbf{Q}$) and that of measurement noises (i.e., $\textbf{R}$), including $\textbf{R}_f$, $\textbf{R}_m$, $\textbf{R}_v$, $\textbf{R}_{\omega}$, $\textbf{R}_{wifi}$, and $\textbf{R}_{mag}$, which represent the noise covariance matrices for accelerometer measurements, magnetometer measurements, velocity/zero velocity updates, zero angular rate updates, WiFi fingerprinting updates, and magnetic fingerprinting updates, respectively. The sign $\gamma(0)$ represents the initial value for the term $\gamma$. The parameters were set at
		 
		 $\textbf{p}^n(0) = \textbf{p}_{wifi}(0); ~\textbf{v}^n(0) = \textbf{0}; ~\bm{\psi}(0)=[\phi(0)~\theta(0)~\psi(0)]^T$
		 
		 $\textbf{x}(0) = [\textbf{0}~\textbf{0}~\textbf{0}~\textbf{0}~\textbf{0}]^T$
		 
		 $\textbf{P}(0) = diag([\textbf{var}_{\textbf{p}^n(0)}~ \textbf{var}_{\textbf{v}^n(0)} ~ \textbf{var}_{\bm{\psi}(0)} ~ \textbf{var}_{\textbf{b}_{g}(0)} ~  \textbf{var}_{\textbf{b}_{a}(0) }])$
 
		$\textbf{Q} = diag( [\textbf{0} ~ \textbf{vrw}^2 ~\textbf{arw}^2 ~ \bm{\epsilon}^2_g ~ \bm{\epsilon}^2_a]) $
		
		$\textbf{R}_f = diag( [\textbf{var}_{f}]); ~ \textbf{R}_m = diag( [\textbf{var}_{m}])$
		
		$\textbf{R}_v = diag( [\textbf{var}_{v}]); ~ \textbf{R}_{\omega} = diag( [\textbf{var}_{\omega}])$
         
         $\textbf{R}_{wifi} = diag( [\textbf{var}_{wifi}]); ~ \textbf{R}_{mag} = diag( [\textbf{var}_{mag}])$  \\
         where $\textbf{p}_{wifi}(0)$ is the initial location from WiFi fingerprinting. $\phi(0)$ and $\theta(0)$ are initial roll and pitch angles calculated by the accelerometer data, while $\psi(0)$ is the initial magnetometer heading. The values of $\textbf{var}_{\textbf{p}^n(0)}$, $\textbf{var}_{\textbf{v}^n(0)}$, $\textbf{var}_{\bm{\psi}(0)} $, $\textbf{var}_{\textbf{b}_{g}(0)}$, and $\textbf{var}_{\textbf{b}_{a}(0) }$ were set at $([ 20~20~20 ]~ m)^2$, $( [ 1~1~1]~ m/s )^2$, $( [ 10~10~90  ] ~deg )^2$, $(  [ 1~1~1  ] ~deg/s)^2$, and $(  [ 0.1~0.1~0.1  ] ~m/s^2 )^2$, respectively. The $\textbf{arw}$, $\textbf{vrw}$, $\bm{\epsilon}_g$, and $\bm{\epsilon}_a$ values were set according to the data sheet of typical low-cost inertial sensors \cite{MPU-6500} as $( [ 0.6~0.6~0.6]~ deg/\sqrt{h} )^2$, $( [ 0.18~0.18~0.18  ] ~m/s/\sqrt{h} )^2$, $(  [ 0.05~0.05~0.05  ] ~deg/s)^2$, and $(  [ 0.01~0.01~0.01  ] ~m/s^2 )^2$, respectively. The values of $\textbf{var}_{f}$, $\textbf{var}_{m}$, $\textbf{var}_{v}$, and $\textbf{var}_{\omega}$ are set at $( [ 2~2~2]~ m/s^2 )^2$, $( [ 0.03~0.03~0.03  ] ~Gauss )^2$, $(  [ 0.3~0.3~0.3  ] ~m/s)^2$, and $(  [ 0.1~0.1~0.1  ] ~deg/s )^2$, respectively. 
         
		Six strategies (i.e., the constant fingerprinting accuracy (CT), SS, SD, WD, MC, and MCM) were used to set values for $\textbf{var}_{wifi}$ and $\textbf{var}_{mag}$. CT represents setting constant values for WiFi and magnetic fingerprinting results, which were set at $([6~6~6]~ m)^2$ and $([5~5~5]~ m)^2$ respectively, according to the statistics of preliminary data. The parameters $\alpha^w_{ss}$ and $\alpha^m_{ss}$ in Equation \eqref{eq-ss-alpha} were set at 0.2 and 50, respectively. The parameters $\sigma^w_{sd}$ and $\sigma^m_{sd}$ in Equation \eqref{eq-sd-alpha} were set at 5 and 1, respectively. The trained $\rho_{ss}$, $\rho_{sd}$, and $\rho_{wd}$ values in Equation \eqref{eq-lin-model-mc} were approximately 0.2, 0.3, and 0.5, respectively. The $\kappa$ and $\kappa_d$ values were set at 5.
		
		 
         \subsection{Fingerprinting Results and Integration with DR}
\begin{figure}
           \centering
           \includegraphics[width=0.45 \textwidth]{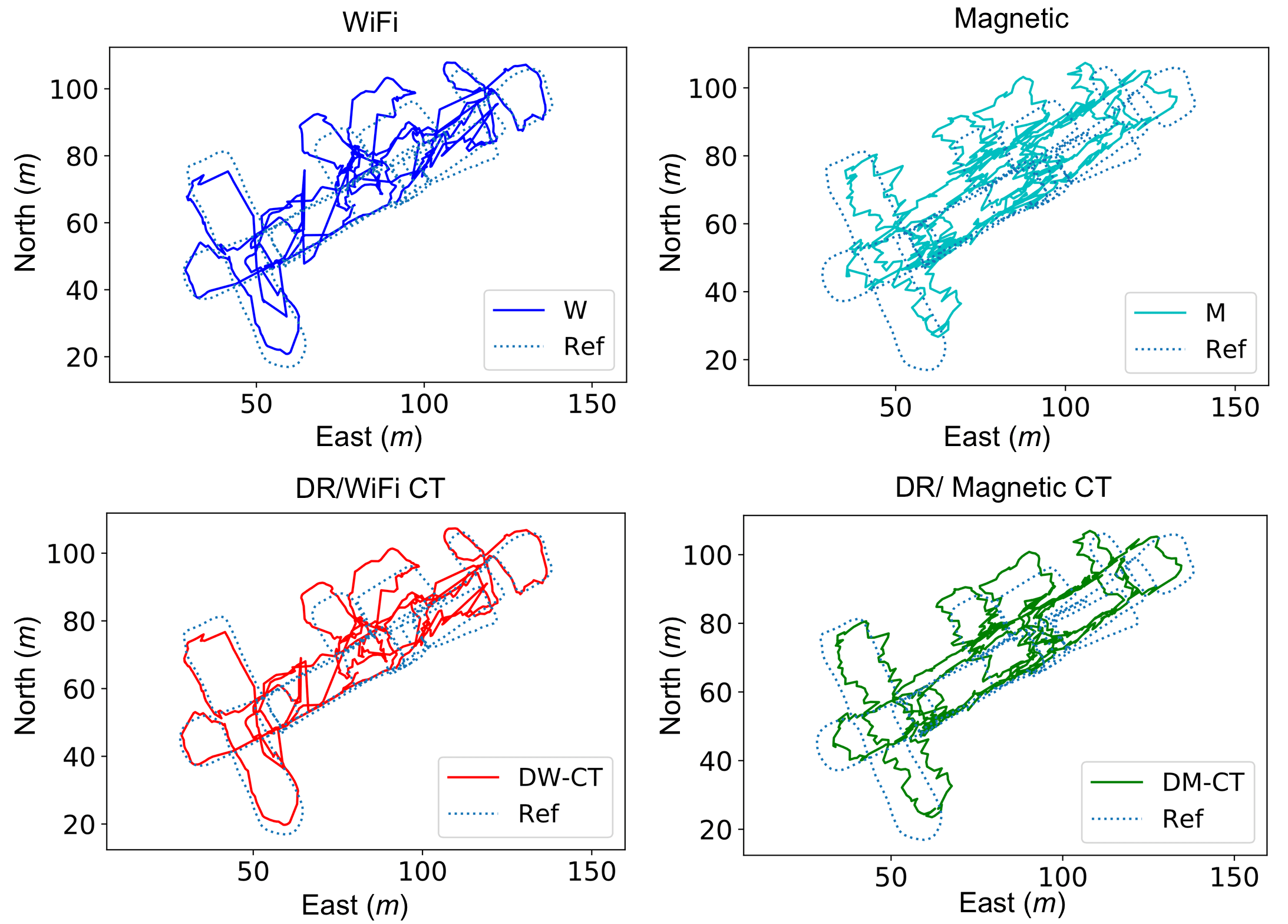}
           \caption{WiFi, magnetic, DR/WiFi-CT, and DR/Magnetic-CT location solutions \cite{LiY2017}}
           \label{fig:wm-loc}
         \end{figure}

\begin{figure}
           \centering
           \includegraphics[width=0.45 \textwidth]{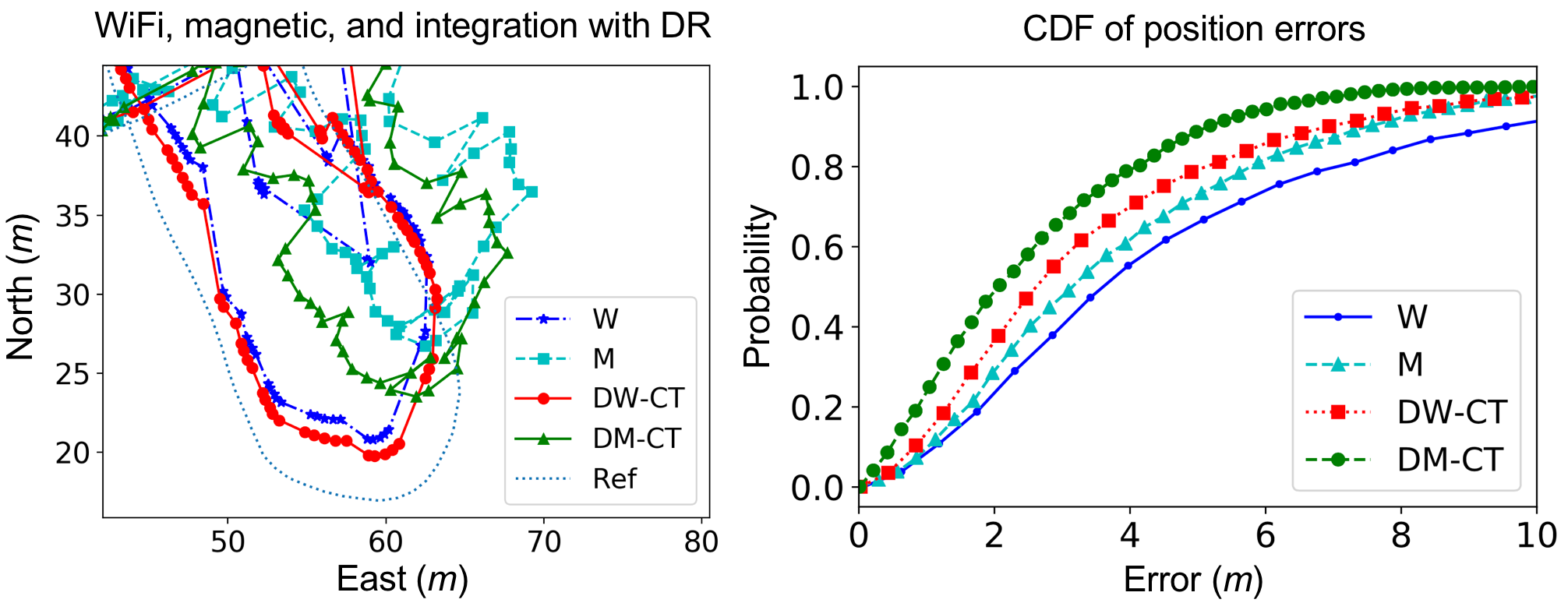}
           \caption{Zoomed-in WiFi, magnetic, DR/WiFi-CT, and DR/Magnetic-CT location solutions \cite{LiY2017} (left) and CDF of location errors (right)} 
           \label{fig:wm-err}
         \end{figure}

         Figure \ref{fig:wm-loc} shows an example WiFi and magnetic localization solution through the method in subsection \ref{sec-fp1}, as well as their integration with DR by using CT in Subsection \ref{sec-test-para}. Figure \ref{fig:wm-err} (left) demonstrates the zoomed-in solutions, while Figure \ref{fig:wm-err} (right) shows the cumulative distribution function (CDF) for localization errors from all test data sets. Table \ref{tab:wm-loc-err} illustrates the error statistics, including the standard deviation (STD), mean value, RMS, the error within which the probability is 80 \% (i.e., the 80 \% error), the error within which the probability is 95 \% (i.e., the 95 \% error), and the maximum value. It is shown that
        
        \begin{itemize}
         \item In Figures \ref{fig:wm-loc} and \ref{fig:wm-err}, both WiFi and magnetic results are generally fitted with the reference trajectories. However, the WiFi solutions suffered from several outliers, while the magnetic solutions suffered from regional errors (i.e., results were biased regionally). These outcomes indicate the importance of accuracy prediction for both WiFi and magnetic fingerprinting, as the accuracy may vary in real time.   
         \item Through integration with DR, the location error RMS values were reduced by 27.1 \% (from 5.9 m to 4.3 m) and 31.1 \% (from 4.5 m to 3.1 m) for WiFi and magnetic fingerprinting, respectively. Meanwhile, the maximum errors were reduced by 27.3 \% (from 27.4 m to 19.9 m) for WiFi and 25.5 \% (from 13.7 m to 10.2 m) for magnetic. These outcomes show the effectiveness of traditional DR/WiFi and DR/Magnetic integration methods \cite{LiY2017} on improving both the accuracy and reliability (i.e., the capability to resist outliers). However, the maximum errors for DR/WiFi integration is still high, which needs to be further reduced. This issue is alleviated in the following subsections.    
         \end{itemize} 
            
         \begin{table}
           \centering
           \begin{tabular}{p{1.5cm} p{0.6cm} p{0.6cm} p{0.6cm} p{0.6cm} p{0.6cm} p{0.6cm} p{0.6cm} }
             \hline
             \textbf{Strategy} & \textbf{STD} ($m$) & \textbf{Mean} ($m$) & \textbf{RMS} ($m$) & \textbf{80\%} ($m$) & \textbf{95\%} ($m$) & \textbf{Max} ($m$) \\ \hline
            	 W  & 3.7 & 4.6 &  5.9  &  7.0 &  12.6 & 27.4     \\ 
             M  & 2.5 & 3.7 &  4.5  &  5.8 &  9.2 & 13.7     \\ 
             DW-CT  & 2.7 & 3.4 &  4.3  &  5.1 &  8.4 & 19.9     \\ 
             DM-CT  & 1.9 & 2.5 &  3.1  &  4.1 &  6.1 & 10.2      \\ \hline
           \end{tabular}
           \caption{Statistics of location errors for WiFi, magnetic, and their integration with DR}
           \label{tab:wm-loc-err}
         \end{table}
         
         \subsection{RSS FAI Results}
		 \subsubsection{WiFi Fingerprinting Accuracy Prediction Results}   
		 \label{sec-res-fai-wifi}      
\begin{figure}
           \centering
           \includegraphics[width=0.45 \textwidth]{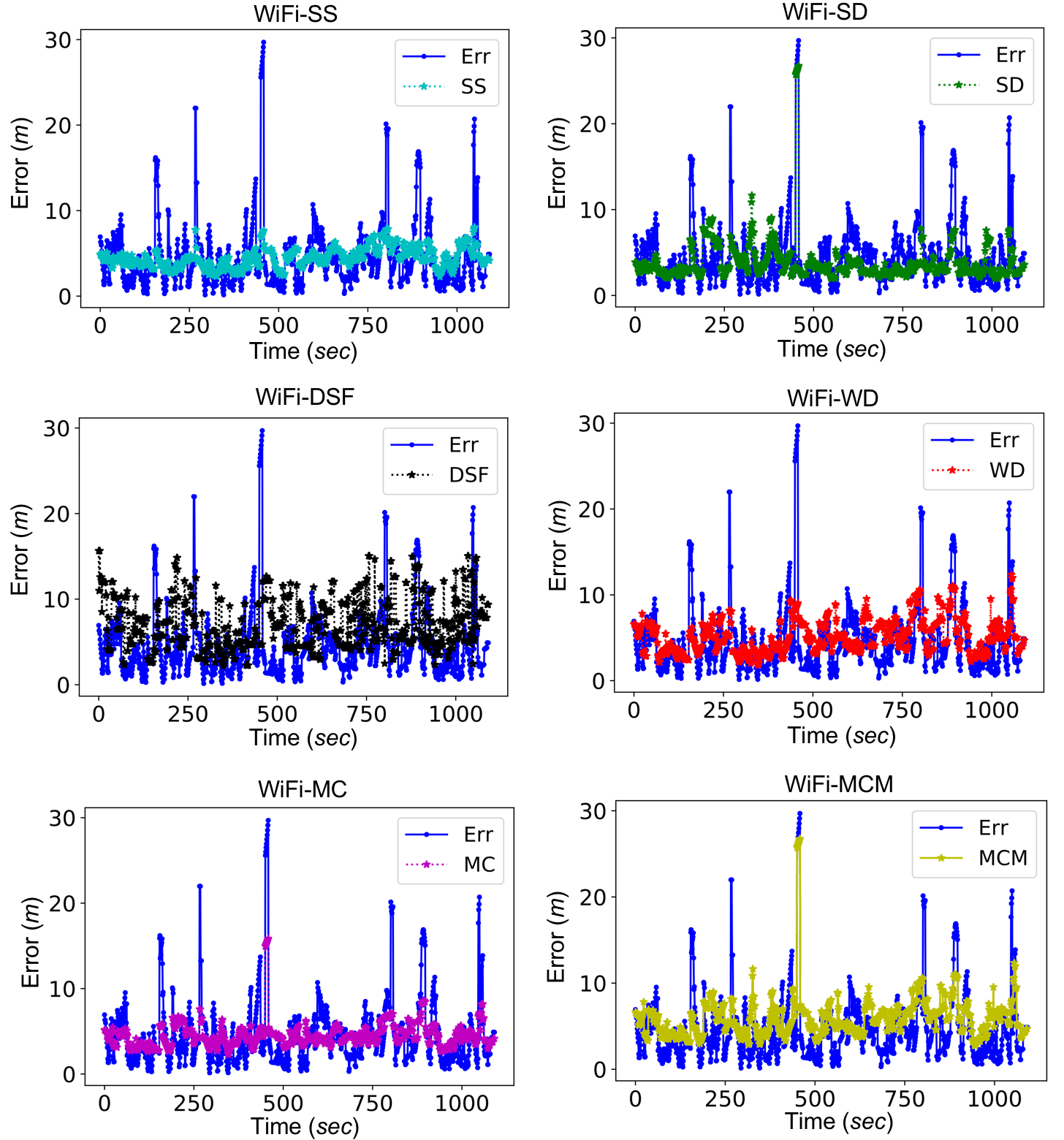}
           \caption{Actual WiFi fingerprinting errors and those predicted by various FAI factors, including SS, SD, DSF \cite{Lemelson2009}, WD, MC, and MCM}
           \label{fig:wifi-err-ss}
         \end{figure}

          To investigate the FAI accuracy, Figure \ref{fig:wifi-err-ss} demonstrates an example of real-time fingerprinting accuracy prediction solutions from SS, SD, DSF \cite{Lemelson2009}, WD, MC, and MCM, as well as the location error time series. It is shown that
         
   	     \begin{itemize}
   	     \item The SS time series had a similar trend with location errors in the long term (e.g., at over 100 s time level). However, SS was not sensitive to short-term (e.g., at 10 s time level) location errors and outliers. This phenomenon can be explained by the limited SS resolution. According to preliminary results, SS was useful for wide-area applications (e.g., building or room detection), but did not have enough resolution for applications within a small area (e.g., moving/static detection). 
         \item SD had predicted several outliers (e.g., at around 450 s) and short-term location errors (e.g., those during 200 to 400 s). However, the SD time series was not clearly correlated with location errors in the long term. One possible reason is that the performance of path-loss model is degraded at some occasions. Thus, SD may be used to detect outliers; however, it is not a reliable indicator for long-term location errors.
		\item DSF predicted the most position errors and outliers. However, the DSF time series was ``noisy". This outcome was consistent with the analysis in Subsection \ref{sec-wd-fai}. Through weighted averaging, the WD time series became smoother and had a more significant correlation with location errors. Time periods that had larger location errors generally had higher WD values. Meanwhile, WD was able to detect outliers. From these outcomes, WD was a more accurate than DSF on WiFi fingerprinting accuracy prediction.
        \item Other than WD, both MC and MCM time series were correlated with location errors. The difference was that the MC time series was smoother, while MCM was more sensitive to outliers. There were occasions at which the predicted MCM value was larger than actual localization accuracy. This phenomenon is acceptable, as discussed in Subsection \ref{sec-mfc}. 
          \end{itemize}
         To evaluate the prediction accuracy, the correlation coefficients $\textit{c}$ between actual location errors and those predicted by FAI factors were calculated by
			\begin{equation}
			\begin{aligned}
			\textit{c}_{\vartheta} = \frac {  \sum_{i=1}^{n_e} { (e_{a,i}- \overline{\textbf{e}_a	}) (e_{{\vartheta},i}- \overline{\textbf{e}_{\vartheta}	})  }   }  
			    {     \sqrt{  \sum_{i=1}^{n_e}{ (e_{a,i} - \overline{\textbf{e}_a})^2 }  }    \sqrt{  \sum_{i=1}^{n_e}{ (e_{{\vartheta},i} - \overline{\textbf{e}_{\vartheta}})^2 }  }        }  
		   \end{aligned}
			\end{equation}		
		  where $\textbf{e}_a$ represents the time series of actual location errors, $\textbf{e}_{\vartheta}$ is the error time series of locations that are predicted by the FAI ${\vartheta}$, where ${\vartheta} \in [[ct],[ss], [sd], [dsf], [wd], [mc], [mcm]]$. $e_{a,i}$ is the $i$-th value in $\textbf{e}_a$. $n_e$ is the number of location errors. The sign $\overline{\bm{\gamma}}$ represents the mean value of $\bm{\gamma}$. Table \ref{tab:cc-wifi} shows the correlation coefficients for multiple FAI factors.
                   
           \begin{itemize}
           \item Excluding CT, WD had the highest score for correlation (0.46), while SS had the lowest (0.21). WD, MC, and MCM had correlation coefficients higher than 0.30. Such results are acceptable when considering there were outliers which are difficult to model and might reduce the correlation coefficient values. 
           \end{itemize}
           \begin{table}
           \centering
           \begin{tabular}{p{0.77cm} p{0.75cm} p{0.75cm} p{0.9cm} p{0.85cm} p{0.85cm} p{1.1cm} }
             \hline
             \textbf{W-CT} & \textbf{W-SS} &  \textbf{W-SD} & \textbf{W-DSF} &\textbf{W-WD} & \textbf{W-MC} & \textbf{W-MCM} \\ \hline
            0.00  & 0.21  & 0.30 & 0.24 &  0.46  &  0.35   & 0.33     \\ \hline
           \end{tabular}
           \caption{Correlation coefficients between actual WiFi fingerprinting errors and those predicted by FAI}
           \label{tab:cc-wifi}
         \end{table}
         
         Table \ref{tab:comp-fai-wifi} summarizes the WiFi FAI factors by their performance on predicting long-term and short-term location errors, outliers, and their computational complexity. The grades S, M, and W represent strong, medium, and weak capabilities, respectively. 
		 \begin{itemize}		 
         \item Generally, WD provided the most accurate prediction on both location errors and outliers, while SS was weak in tracking short-term errors and outliers, and SD was relatively weak in indicating long-term errors. MC was strong in predicting location errors but missed several outliers, while MCM predicted the majority of location errors and outliers.
		\item On the other hand, the time complexity of  WD, MC, and MCM are $O(n_{p} n_w)$, which are higher than that of CT ($O(1)$), SS ($O(n_w)$), and SD ($O(n_w)$). $n_w$ represents the dimension of RSS feature vector and $n_p$ denotes the number of RPs in the database.   
		\end{itemize}  
         \begin{table}
           \centering
           \begin{tabular}{p{1.2cm} p{1.3cm} p{1.35cm} p{1.25cm} p{1.25cm} }
             \hline
             \textbf{Strategy} & \textbf{Long-term} & \textbf{Short-term} & \textbf{Outlier} & \textbf{Complexity}  \\ \hline
             W-CT & W*  & W & W  &  $O(1)$    \\ 
             W-SS & M  & W  & W  & $O(n_w)$     \\ 
             W-SD & W  & M  & M  & $O(n_w)$      \\ 
             W-WD & S  & S  & M  & $O(n_{p} n_w)$      \\ 
             W-MC & S  & S  & M  & $O(n_{p} n_w)$       \\ 
             W-MCM & S  & S  & S  & $O(n_{p} n_w)$     \\ \hline
           \end{tabular}
            \begin{tablenotes}
        		\item[*] * S-strong; M-medium; W-weak; $n_w$-dimension of RSS feature vector; $n_p$-number of RPs in database 
     		\end{tablenotes}
           \caption{ Performance of WiFi FAI factors  }
           \label{tab:comp-fai-wifi}
         \end{table}
        
         \subsubsection{FAI-enhanced DR/WiFi Integration Results}           
\begin{figure}
           \centering
           \includegraphics[width=0.45 \textwidth]{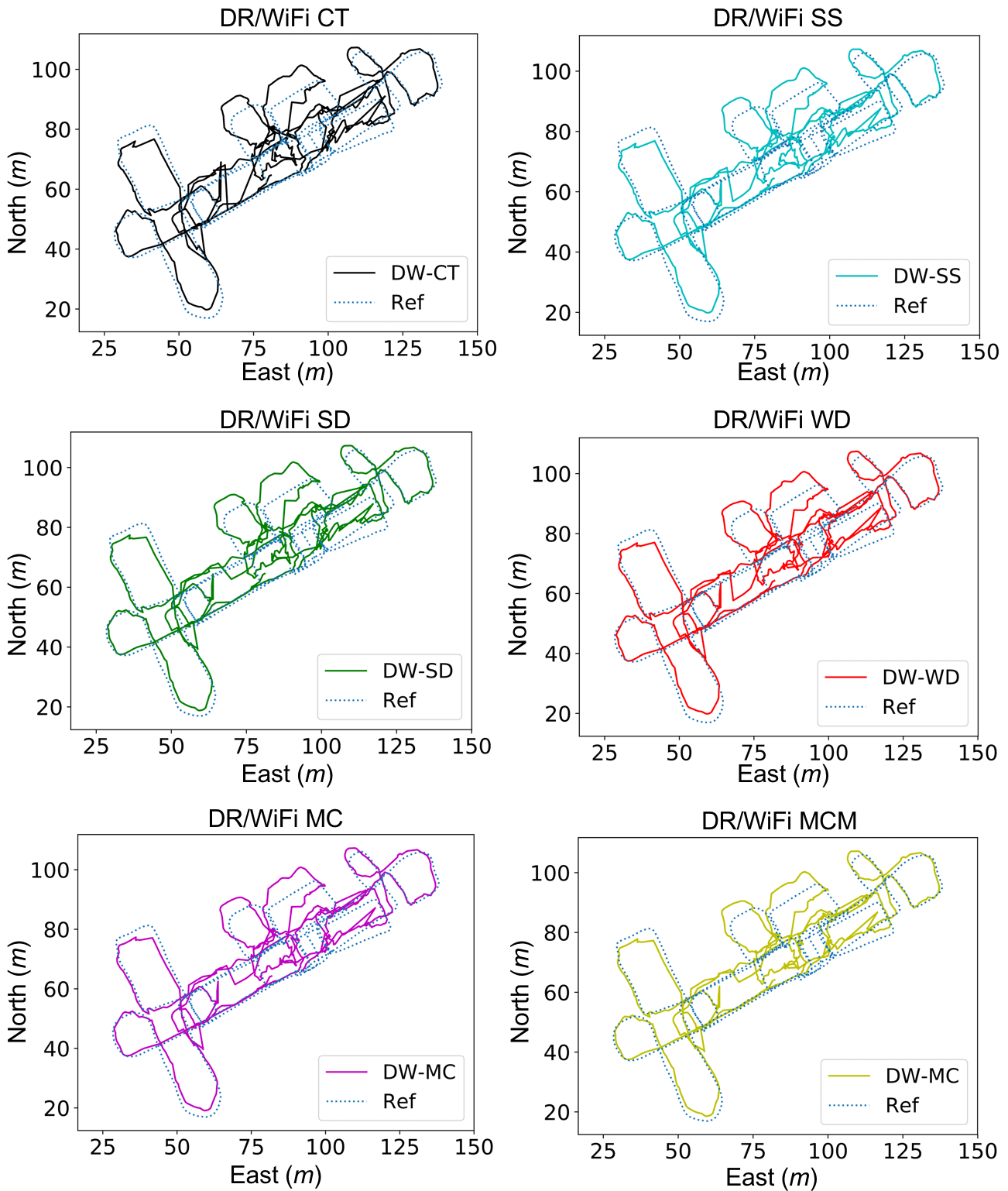}
           \caption{DR/WiFi solutions with various FAI strategies}
           \label{fig:dw-loc}
         \end{figure}

\begin{figure}
           \centering
           \includegraphics[width=0.45 \textwidth]{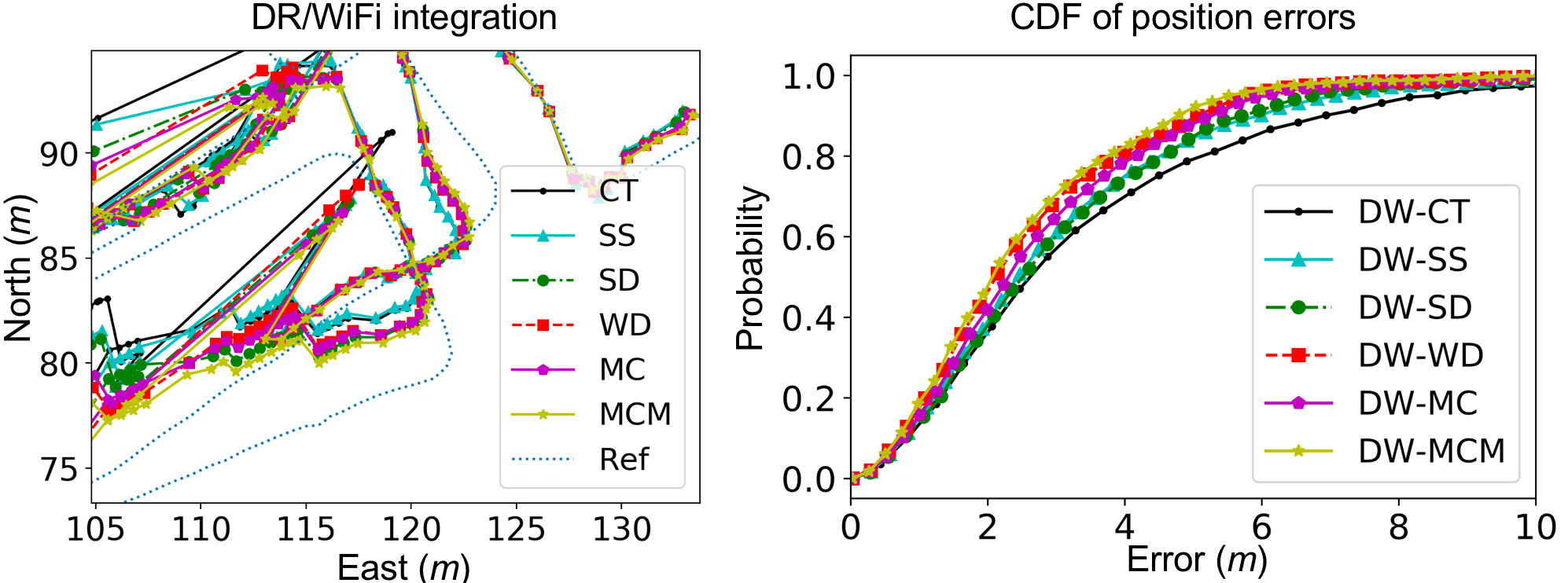}
           \caption{Zoomed-in DR/WiFi location solutions (left) and CDF of location errors (right) with various FAI strategies}
           \label{fig:dw-err}
         \end{figure}

          Figure \ref{fig:dw-loc} shows a set of DR/WiFi location solutions that used various strategies for setting WiFi position measurement noises, while Figure \ref{fig:dw-err} (left) demonstrates the zoomed-in solutions and Figure \ref{fig:dw-err} (right) shows the CDF of location errors from all test data. Table \ref{tab:dw-loc-err} illustrates the corresponding error statistics. Figures \ref{fig:dw-loc}, \ref{fig:dw-err} and Table \ref{tab:dw-loc-err} indicate that
          \begin{table}
           \centering
           \begin{tabular}{p{1.5cm} p{0.6cm} p{0.6cm} p{0.6cm} p{0.6cm} p{0.6cm} p{0.6cm} p{0.6cm} }
             \hline
             \textbf{Strategy} & \textbf{STD} ($m$) & \textbf{Mean} ($m$) & \textbf{RMS} ($m$) & \textbf{80\%} ($m$) & \textbf{95\%} ($m$) & \textbf{Max} ($m$) \\ \hline
            	 DW-CT & 2.7  & 3.4  & 4.3  & 5.1  & 8.4  & 19.9     \\ 
             DW-SS & 2.1  & 3.0  & 3.6  & 4.5  & 7.1  & 13.3     \\ 
             DW-SD & 2.0  & 3.0  & 3.7  & 4.6  & 6.7  & 12.7     \\ 
             DW-WD & 1.8  & 2.6  & 3.1  & 3.9  & 5.8  & 13.8     \\ 
             DW-MC & 1.8  & 2.7  & 3.3  & 4.2  & 6.0  & 12.0     \\ 
             DW-MCM & 1.7  & 2.5  & 3.0  & 3.8  & 5.5  & 11.7    \\ \hline
           \end{tabular}
           \caption{Statistics of DR/WiFi integrated location errors}
           \label{tab:dw-loc-err}
         \end{table}   
                 
         \begin{itemize}
         \item The use of SS, SD, WD, MC, and MCM had improved DR/WiFi integrated location accuracy (RMS errors) by 16.3 \%, 14.0 \%, 27.9 \%, 23.3 \%, and 30.2 \%, respectively, and had reduced the maximum errors by 33.2 \%, 36.2 \%, 30.6 \%, 39.7 \%, and 41.2 \%, respectively. These outcomes illustrated the effectiveness of FAI on enhancing DR/WiFi integrated location.          
         \item DW-WD provided higher prediction accuracy than DW-SS and DW-SD but suffered from the largest maximum errors. This outcome was consistent with the FAI prediction solutions, as WD had an advantage in predicting long-term and short-term errors, instead of outliers. 
         \item The performance of DW-MC was a trade-off among DW-SS, DW-SD, and DW-WD. Thus, DW-MC was less accurate than DW-WD, but stronger in reducing maximum errors. DW-MCM also provided accurate and reliable location solutions. Therefore, although SS and SD did not provide predictions that were as accurate as WD, they may still be combined with WD to further enhance fingerprinting accuracy prediction. 
         \end{itemize}
         
         \subsection{Magnetic FAI Results} 
         \subsubsection{Magnetic Fingerprinting Accuracy Prediction Results}    
         \label{sec-res-fai-mag}      
\begin{figure}
           \centering
           \includegraphics[width=0.45 \textwidth]{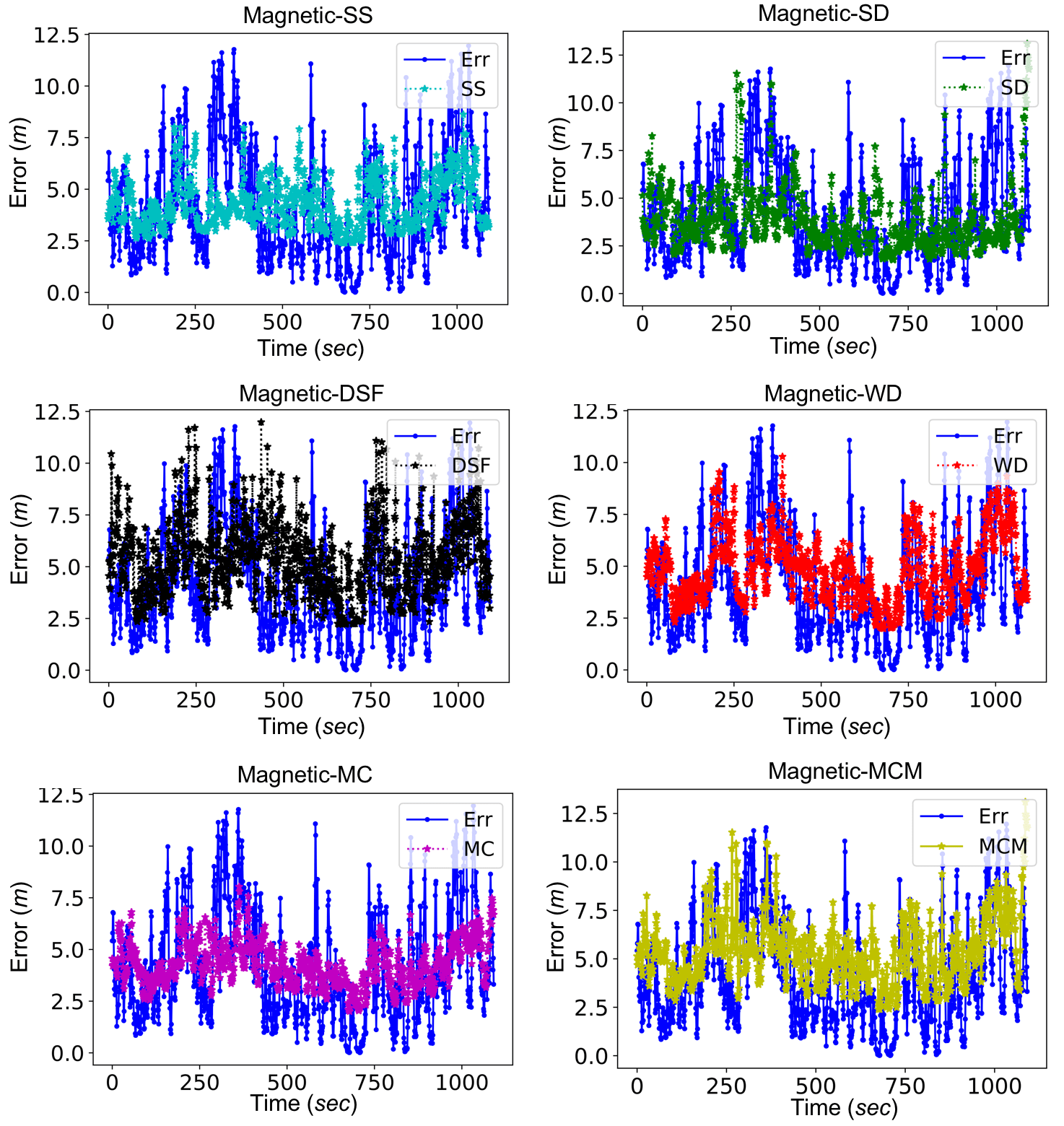}
           \caption{Actual magnetic fingerprinting errors and those predicted by various FAI factors}
           \label{fig:mag-err-ss}
         \end{figure}

         Figure \ref{fig:mag-err-ss} demonstrates the real-time magnetic fingerprinting errors predicted by SS, SD, DSF, WD, MC, and MCM, respectively. It is shown that 
         \begin{itemize}
         \item Compared with WiFi, the magnetic fingerprinting errors had more significant long-term trends because magnetic data was more susceptible to regional localization errors. 
         \item WD was capable to track most long-term errors, while SS and SD tracked part of them (e.g., 800 to 1000 s for SS, and 200 to 400 s for SD) but missed others. 
         \item Both MC and MCM predicted the majority of location errors. Compared to MC, MCM had a larger fluctuation range. 
         \end{itemize}
                 
         Table \ref{tab:cc-mag} demonstrates the correlation coefficients between actual magnetic fingerprinting errors and those predicted by FAI factors. It can be found that 
         \begin{itemize}
         \item Compared to WiFi fingerprinting, the magnetic correlation coefficients were higher (0.60) for WD but lower for SS (0.12) and SD (0.18). The higher WD value was partly because there were less outliers in magnetic fingerprinting results. The low values for SS and SD indicated the difficulty in predicting magnetic fingerprinting errors from the signal strength or geometric levels, as magnetic data had a lower dimension. 
         \item After combination, MC and MCM had similar correlation coefficient values (i.e., 0.51 and 0.45). Similar to WiFi, WD, MCM, and MC were the three FAI factors that had the highest correlation coefficients for magnetic fingerprinting. 
         \end{itemize} 
         \begin{table}
           \centering
           \begin{tabular}{p{0.77cm} p{0.75cm} p{0.75cm} p{0.9cm} p{0.85cm} p{0.85cm} p{1.12cm} }
             \hline
             \textbf{M-CT} & \textbf{M-SS} & \textbf{M-SD} & \textbf{M-DSF} & \textbf{M-WD} & \textbf{M-MC} & \textbf{M-MCM} \\ \hline
            	0.00  & 0.12 & 0.19 &  0.37 &  0.60  &  0.51   & 0.45     \\ \hline
           \end{tabular}
           \caption{Correlation coefficients between actual magnetic fingerprinting errors and those predicted by FAI}
           \label{tab:cc-mag}
         \end{table}   
                 
       Table \ref{tab:comp-fai-mag} summarizes the performances of magnetic FAI factors on predicting location errors, outliers, and their computational complexity. A main difference to the WiFi results is that SD becomes weaker in detecting outliers in magnetic data, while WD becomes stronger in detection outliers. 
         \begin{table}
           \centering
           \begin{tabular}{p{1.2cm} p{1.3cm} p{1.35cm} p{1.25cm} p{1.25cm} }
             \hline
             \textbf{Strategy} & \textbf{Long-term} & \textbf{Short-term} & \textbf{Outlier} & \textbf{Complexity}  \\ \hline
            	 M-CT & W*  & W  & W  & $O(1)$    \\ 
             M-SS & M  & W  & W  & $O(n_m)$     \\ 
             M-SD & W  & M  & W  & $O(n_m)$      \\ 
             M-WD & S  & S  & S  & $O(n_{p} n_m)$     \\ 
             M-MC & S  & S  & M  & $O(n_{p} n_m)$       \\ 
             M-MCM & S  & S  & S  & $O(n_{p} n_m)$    \\ \hline
           \end{tabular}
            \begin{tablenotes}
        		\item[*] * S-strong; M-medium; W-weak; $n_m$-dimension of magnetic feature vector; $n_p$-number of RPs in database
     		\end{tablenotes}
           \caption{ Performance of magnetic FAI factors  }
           \label{tab:comp-fai-mag}
         \end{table}

          \subsubsection{FAI-enhanced DR/Magnetic Integration Results}   
\begin{figure}
           \centering
           \includegraphics[width=0.45 \textwidth]{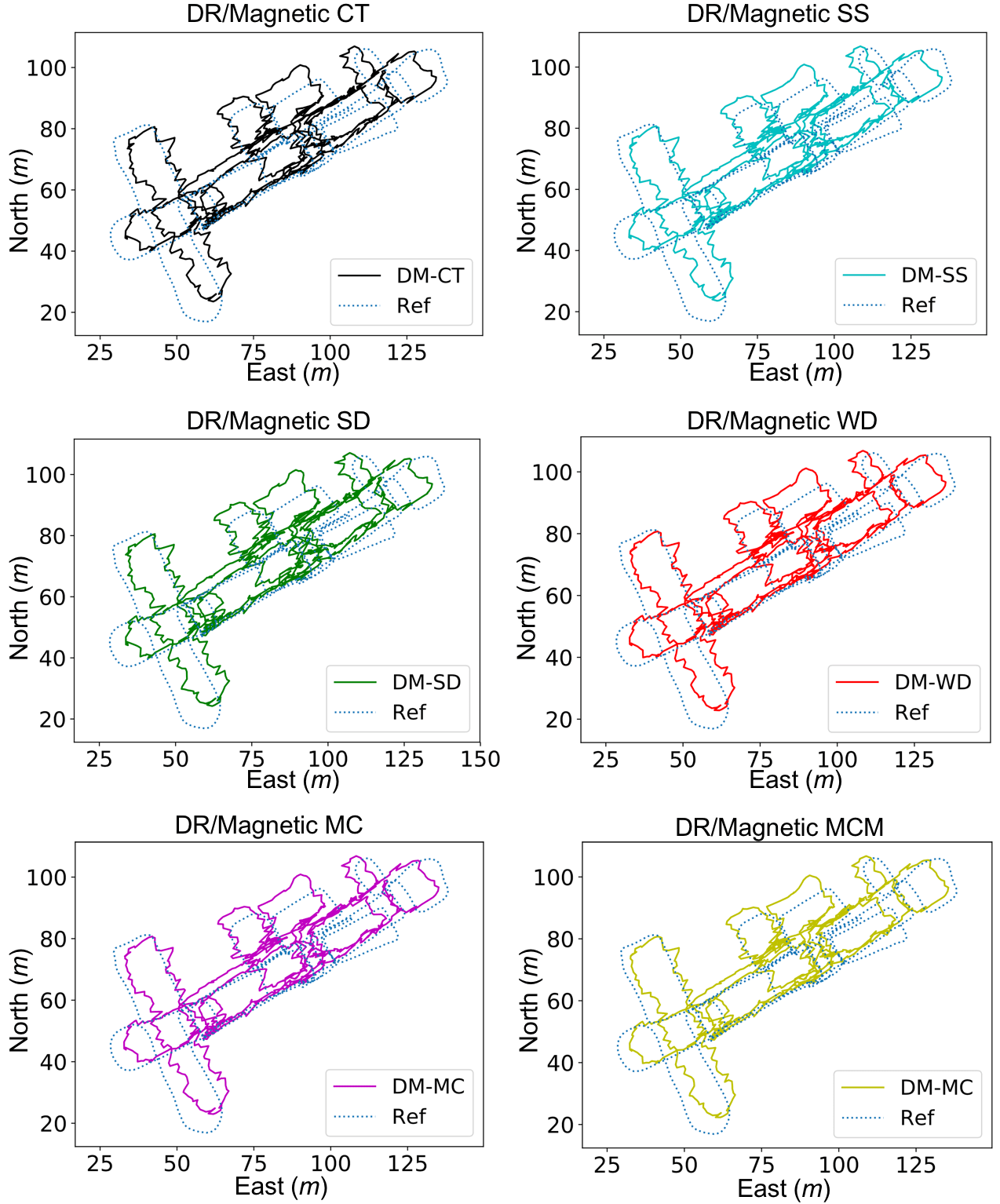}
           \caption{DR/Magnetic solutions with various FAI strategies}
           \label{fig:dm-loc}
           \end{figure}

\begin{figure}
           \centering
           \includegraphics[width=0.45 \textwidth]{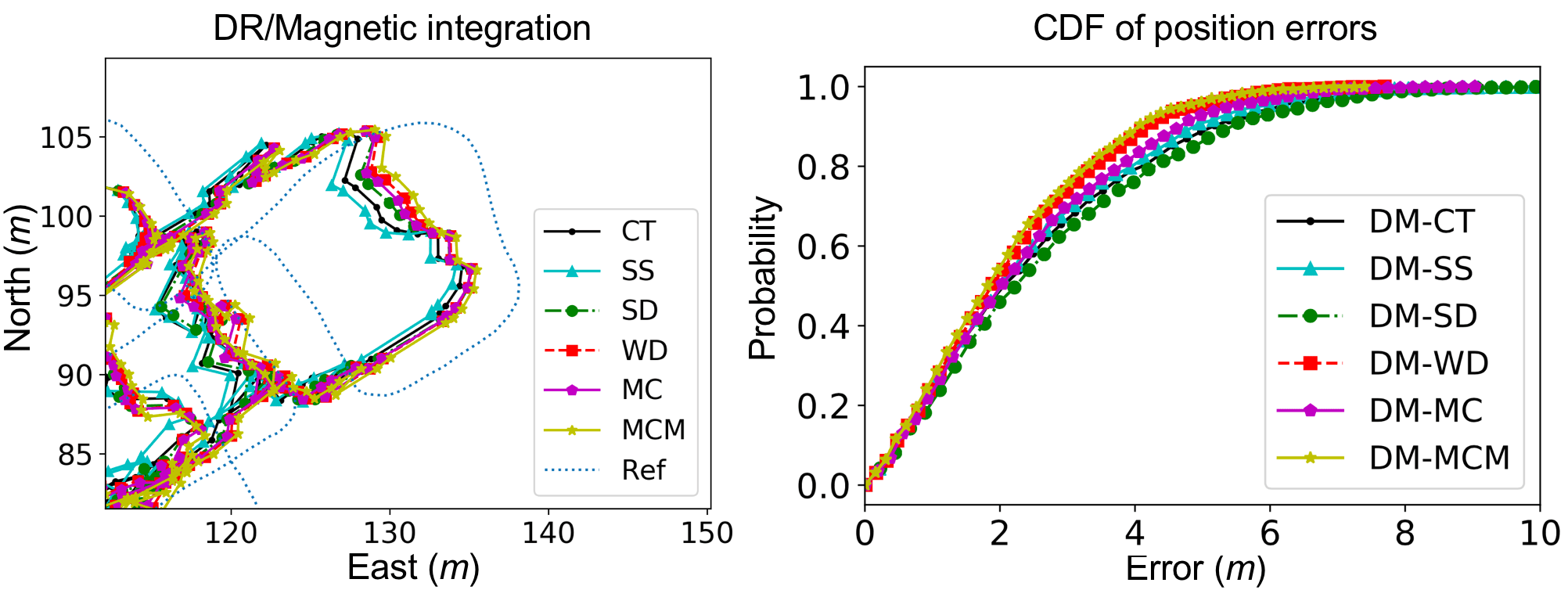}
           \caption{Zoomed-in DR/Magnetic location solutions (left) and CDF of location errors (right) with various FAI strategies}
           \label{fig:dm-err}
         \end{figure}

          Figure \ref{fig:dm-loc} shows a set of DR/Magnetic localization solutions that used various strategies for setting position measurement noises, while Figure \ref{fig:dm-err} (left) demonstrates the zoomed-in solutions and Figure \ref{fig:dm-err} (right) shows the CDF of location errors from all test data. Table \ref{tab:dm-loc-err} illustrates the corresponding error statistics. Figures \ref{fig:dm-loc}, \ref{fig:dm-err} and Table \ref{tab:dm-loc-err} show that 
          \begin{table}
           \centering
           \begin{tabular}{p{1.5cm} p{0.6cm} p{0.6cm} p{0.6cm} p{0.6cm} p{0.6cm} p{0.6cm} p{0.6cm} }
             \hline
             \textbf{Strategy} & \textbf{STD} ($m$) & \textbf{Mean} ($m$) & \textbf{RMS} ($m$) & \textbf{80\%} ($m$) & \textbf{95\%} ($m$) & \textbf{Max} ($m$) \\ \hline
            	 DM-CT &  1.9   & 2.5   & 3.1   & 4.1   & 6.1   & 10.2     \\ 
             DM-SS &  1.8   & 2.4   & 3.0   & 3.9   & 5.8   & 10.1     \\ 
             DM-SD &  1.9   & 2.7   & 3.3   & 4.3   & 6.5   & 10.8     \\ 
             DM-WD &  1.4   & 2.1   & 2.6   & 3.4   & 4.9   & 7.7     \\ 
             DM-MC &  1.6   & 2.3   & 2.9   & 3.8   & 5.4   & 9.0     \\ 
             DM-MCM &  1.4   & 2.1   & 2.5   & 3.3   & 4.7   & 7.4   \\ \hline
           \end{tabular}
           \caption{Statistics of DR/Magnetic location errors}
           \label{tab:dm-loc-err}
         \end{table}
         
         \begin{itemize} 
         \item DM-SS provided similar localization accuracy to DM-CT, while DM-SD degraded the solutions by 6.1 \%. This outcome indicates the importance of using a proper FAI strategy. 
         \item Similar to the DR/WiFi case, DM-WD, DM-MC, and DM-MCM had improved location accuracy by 16.1 \%, 6.5 \%, and 19.4 \%, and reduced the maximum errors by 24.5 \%, 15.7 \% and 28.4 \%, respectively. 
         \item In contrast to DW-WD, DM-WD provided lower maximum errors because magnetic fingerprinting solutions had less outliers. 
         \end{itemize}
         
         \subsection{FAI-enhanced DR/WiFi/Magnetic Integration Results}   
\begin{figure}
           \centering
           \includegraphics[width=0.45 \textwidth]{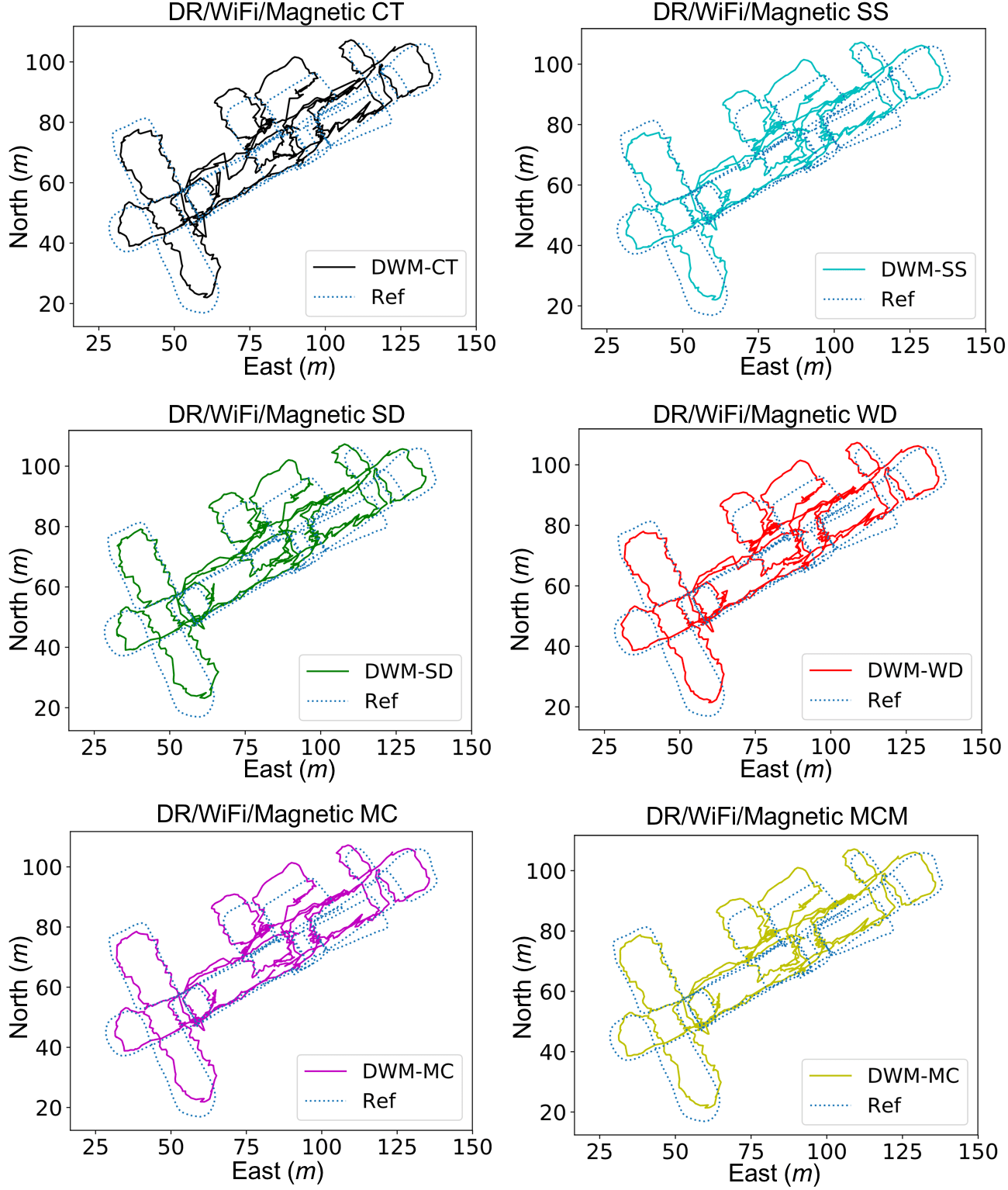}
           \caption{DR/WiFi/Magnetic solutions with various FAI strategies}
           \label{fig:dwm-loc}
         \end{figure}

\begin{figure}
           \centering
           \includegraphics[width=0.45 \textwidth]{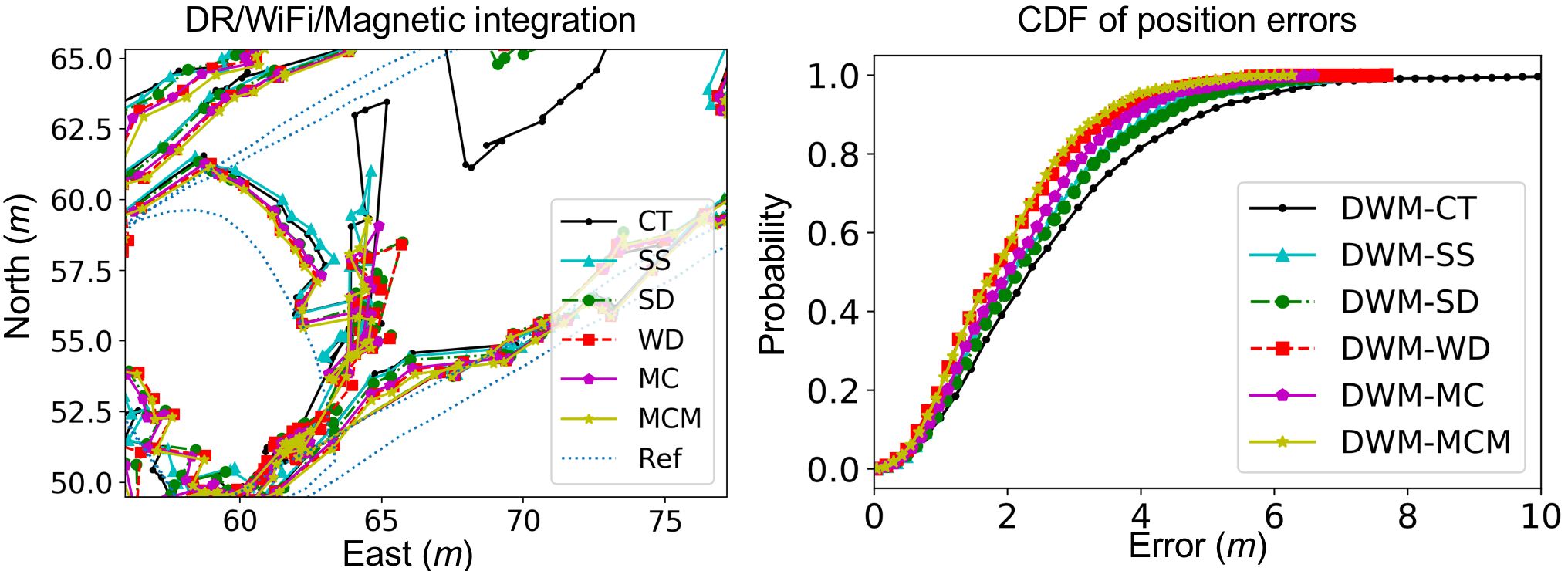}
           \caption{Zoomed-in DR/WiFi/Magnetic location solutions (left) and CDF of location errors (right) with various FAI strategies}
           \label{fig:dwm-err}
         \end{figure}

          Figure \ref{fig:dwm-loc} shows a set of DR/WiFi/Magnetic solutions that used various strategies for position measurement noises, while Figure \ref{fig:dwm-err} (left) demonstrates the zoomed-in solutions and Figure \ref{fig:dwm-err} (right) shows the CDF of localization errors from all test data. Table \ref{tab:dwm-loc-err} illustrates the error statistics. It is shown that
          \begin{table}
           \centering
           \begin{tabular}{p{1.5cm} p{0.6cm} p{0.6cm} p{0.6cm} p{0.6cm} p{0.6cm} p{0.6cm} p{0.6cm} }
             \hline
             \textbf{Strategy} & \textbf{STD} ($m$) & \textbf{Mean} ($m$) & \textbf{RMS} ($m$) & \textbf{80\%} ($m$) & \textbf{95\%} ($m$) & \textbf{Max} ($m$) \\ \hline
            	 DWM-CT   & 1.7  & 2.7  & 3.1  & 3.9  & 5.9  & 11.3     \\ 
             DWM-SS   & 1.3  & 2.3  & 2.7  & 3.4  & 4.8  & 7.2     \\ 
             DWM-SD   & 1.4  & 2.4  & 2.8  & 3.5  & 5.0  & 7.3     \\ 
             DWM-WD   & 1.2  & 2.0  & 2.3  & 2.9  & 4.1  & 7.7     \\ 
             DWM-MC   & 1.2  & 2.2  & 2.5  & 3.1  & 4.5  & 6.6     \\ 
             DWM-MCM   & 1.1  & 1.9  & 2.2  & 2.8  & 3.9  & 6.3    \\ \hline
           \end{tabular}
           \caption{Statistics of DR/WiFi/Magnetic integrated location errors}
           \label{tab:dwm-loc-err}
         \end{table}    
         
        \begin{itemize}
        \item Similar to the existing works, DR/WiFi/Magnetic integrated location solutions were more accurate than those from DR/WiFi and DR/Magnetic integration.       
        \item The introduction of FAI further improved the DR/WiFi/Magnetic solution. To be specific, DWM-SS, DWM-SD, DWM-WD, DWM-MC, and DWM-MCM had improved location accuracy by 12.9 \%, 9.7 \%, 25.8 \%, 19.4 \%, and 29.0 \%, and reduced the maximum errors by 36.3 \%, 35.4 \%, 31.9 \%, 41.6 \% and 44.2 \%, respectively, compared to the DWM-CT case.    
        \item MCM, WD, and MC were the top three FAI strategies that provided the highest DR/WiFi/Magnetic integrated localization accuracy, while MCM and MC were the strongest FAI in resisting outliers.  
        \end{itemize}
        
        Table \ref{tab:dw-top-three} illustrates the top three FAI strategies in terms of accuracy and reliability based on comprehensive analysis on DR/WiFi, DR/Magnetic, and DR/WiFi/Magnetic integrated solutions. MCM, WD, and MC were three most effective FAI strategies to the tests. It is notable that the purpose of Table \ref{tab:dw-top-three} is to provide a summary of test results, instead of drawing a conclusion that MCM is a superior FAI factor than others. The reason is that the performance of FAI factors may vary according to real-world application scenarios. However, the proposed crowdsourcing-based localization method can be straightforwardly used in other scenarios and applications. Meanwhile, the outcomes from this paper provide a reference for using FAI to predict fingerprinting accuracy and enhancing multi-sensor integrated localization.     
        \begin{table}
           \centering
           \begin{tabular}{p{2.0cm} p{0.8cm} p{0.8cm} p{0.8cm}  }
             \hline
             \textbf{Term} & \textbf{DW} & \textbf{DM} & \textbf{DWM}  \\ \hline
            	 Accuracy 1\textsuperscript{st}  & MCM  &  MCM  &    MCM   \\ 
             Accuracy 2\textsuperscript{nd}  & WD   &  WD   &    WD \\ 
             Accuracy 3\textsuperscript{rd}  & MC   &  MC   &    MC \\ 
             Reliability 1\textsuperscript{st} &  MCM  &  MCM & MCM    \\ 
             Reliability 2\textsuperscript{nd} &  MC   &  MC  &  MC   \\ 
             Reliability 3\textsuperscript{rd} & SD    &  WD  &  SS \\ \hline
           \end{tabular}
           \caption{FAI factors with highest prediction accuracy and reliability}
           \label{tab:dw-top-three}
         \end{table}
         
         \ifCLASSOPTIONcaptionsoff
         \newpage
         \fi

         \section{Conclusions}
         \label{sec-conclusionsl-work}
         The unpredictability of fingerprinting accuracy has greatly limited the localization-enhanced IoT applications. Till today, most existing indoor localization works focus on improving localization techniques, while there is a lack of investigation of indicator metric regarding the localization accuracy, especially fingerprinting accuracy. Therefore, this paper introduces the fingerprinting accuracy indicator (FAI) into multi-sensor integrated localization to fill this gap. FAI factors from the signal strength, geometry, and database levels and their combinations have been proposed and evaluated. It is found that a) the proposed weighted distance between similar fingerprints (weighted DSF) based FAI was effective in predicting both location errors and outliers. b) The geometry based FAI was stronger in predicting short-term errors and outliers, while c) the signal strength based FAI was stronger in predicting long-term location errors. Due to such complementary characteristics, FAI combination strategies were adopted to provide accurate and robust fingerprinting accuracy prediction.
         
         Furthermore, this paper proposes an enhancement to the crowdsourcing-based localization method, which does not involve user intervention or parameter tuning. Crowdsourced sensor data is used for updating both wireless and magnetic databases simultaneously while localizing. Additionally, the FAI-enhanced dead-reckoning/wireless/magnetic integrated extended Kalman filter (EKF) is used for robust localization. By using off-the-shelf inertial sensors in consumer devices and existing WiFi access points (which provided location error root mean square (RMS) of 5.9 m and maximum error of 27.4 m) and the indoor magnetic field (which provided location error RMS of 4.5 m and maximum error of 13.1 m), the FAI-enhanced EKF provided localization solutions that had error RMS of 2.2 m and maximum error of 6.3 m. These values were respectively 29.0 \% and 44.2 \% more accurate than those from EKF without FAI. These outcomes indicate the effectiveness of the proposed FAI on improving both accuracy and reliability of multi-sensor integrated localization.   



\end{document}